\documentclass[prd,nopacs,nokeys]{revtex4}
\usepackage{amsfonts}
\usepackage{amsmath}
\usepackage{amssymb}
\usepackage{bbm}
\usepackage[utf8]{inputenc}
\usepackage[caption=false]{subfig}
\usepackage{tensor}
\usepackage{slashed}
\usepackage[centertableaux ]{ytableau}
\usepackage{hyperref}
\usepackage{natbib}
\usepackage{enumerate}
\usepackage{braket,mleftright}
\usepackage{graphicx}
\usepackage{cleveref}

\providecommand{\U}[1]{\protect\rule{.1in}{.1in}}

\renewcommand{\vec}[1]{\mathbf{#1}} 

\begin{document}

\title{Spin-one \textit{matter} fields.}
\author{M. Napsuciale $^{(1)}$, S. Rodr\'{\i}guez $^{(2)}$, Rodolfo
Ferro-Hern\'{a}ndez $^{(1)}$, Selim G\'omez-\'Avila $^{(1)}$}

\begin{abstract}
{\ Spin-one matter fields are relevant both for the description of hadronic
states and as potential extensions of the Standard Model. In this work we
present a formalism for the description of massive spin-one fields
transforming in the $(1,0)\oplus(0,1)$ representation of the Lorentz group,
based on the covariant projection onto parity eigenspaces and Poincar\'e
orbits. The formalism yields a constrained dynamics. We solve the
constraints and perform the canonical 
quantization accordingly. This formulation uses the recent construction of a 
parity-based covariant basis for matrix operators acting on the $(j,0)\oplus(0,j) $ 
representations. The algebraic properties of the covariant basis play an important 
role in solving the constraints and allowing the canonical quantization of the
theory. We study the chiral structure of the theory and conclude that it
is not chirally symmetric in the massless limit, hence it is not possible 
to have chiral gauge interactions. However, spin-one matter fields 
can have vector gauge interactions. Also, the dimension of the field makes 
self-interactions naively renormalizable. Using the covariant basis, we 
classify all possible self-interaction terms. }
\end{abstract}

\maketitle

\address{$^{1}$ Departamento de F\'{\i}sica, Universidad de
  Guanajuato, Lomas del Bosque 103, Fraccionamiento Lomas del
  Campestre, 37150, Le\'{o}n, Guanajuato, M\'{e}xico.} 
\address{$^{2}$ Facultad de Ciencias F\'isico-Matem\'aticas,
  Universidad Aut\'onoma de Coahuila, Edificio A, Unidad
  Camporredondo, 25000, Saltillo Coahuila M\'exico}



\section{Introduction.}


States transforming in the $(1,0)\oplus(0,1)$ representation have been
shown to be appropriate for the description of low-energy interactions
of the low-lying nonets of vector and axial-vector mesons
\cite{Ecker:1988te}. The corresponding fields are written in tensor
language (an antisymmetric second-rank tensor field is used to
describe spin-one mesons) and the effective theory known as Resonance
Chiral Perturbation Theory ($R\chi PT$) involves a nonlinear
realization of chiral symmetry. Also, possible effects of spin-one
matter particles described by tensor fields in physics beyond the
standard model have been proposed in \cite{Chizhov:2011zz}.

On the other hand, many alternatives for physics beyond the standard model
have been proposed and although the first results of the Large Hadron
Collider (LHC) showed no evidence of any of these possibilities up to
energies of the order of $1.5 ~TeV$ \cite{Aad20121,Chatrchyan201230}, \cite%
{Chatrchyan:2013gia}, \cite{Chatrchyan:2013xva}, \cite{Chatrchyan:2013oca},%
\cite{Khachatryan:2014dka}, \cite{Khachatryan:2014xja}, \cite%
{Khachatryan:2014aka}, \cite{Shifman:2012na}, recently a series of excess of
events in several searches of new spin-one bosons at the level of 2-3
standard deviations, point to the possible existence of new spin-one
resonances close to $2 ~TeV$ \cite{Aad:2015owa}. The simplest possibility
for these resonances is some realization of the left-right symmetric models
and the first possible explanations of the excess of events following this
route have been already proposed in \cite{Dobrescu:2015yba}, \cite%
{Dev:2015pga}. An alternative to the understanding these events would be
offered by spin-one matter fields. Indeed, it is intriguing that the
standard model and most of the proposed non-supersymmetric extensions use
only the $(0,0)$, $(1/2,0)$, $(0,1/2)$ and $(1/2,1/2)$ representations of
the Homogeneous Lorentz Group (HLG). The consistent formulation of a theory
involving fields transforming in the chiral $(1,0)$ and $(0,1)$
representations of the HLG would certainly enlarge the possibilities for
beyond the standard model theories.

Recently, an algorithm for the construction of a covariant basis for the
matrix operators acting on the $(j,0)\oplus(0,j)$ representation space was
put forth in Ref. \cite{Gomez-Avila:2013qaa}. This construction is based on
the covariant properties of the parity operator, and the explicit form of
the covariant matrices is given for $j=1/2, 1, 3/2$. For $j=1/2$ the
covariant basis reproduces the conventional basis acting on Dirac space, and
the Dirac equation is recovered as the covariant projection onto parity
eigenspaces. This alternative view of the Dirac equation, and the fact that
the covariant basis for $(1,0)\oplus(0,1)$ has been already constructed in 
\cite{Gomez-Avila:2013qaa}, leads us to explore the $j=1$ generalization of
the structure of the Dirac theory. Since a chirality operator appears in a
natural way in the covariante basis, chiral states can be constructed
directly. This allows us to study alternatives for the formulation of chiral
effective theories for hadrons using the Dirac-like theory for fields
transforming in the $(1,0)\oplus (0,1)$ representation of the HLG.

In this work, we propose a theory for massive spin-one matter fields which is a
direct generalization to $j=1$ of the structure of the Dirac theory for
fermions. The formalism is based on the simultaneous projection onto
invariant parity subspaces and appropriate Poincar\'e orbit. The formalism
yields a constrained dynamics with second class constraints. We work out
these constraints in the classical field theory, and show that sensible
results are obtained upon quantization once we use the specific algebraic
properties of the covariant basis. We study the chiral structure and
classify the naively renormalizable self-interactions of the spin-one matter fields.

Our paper is organized as follows. In the next section we introduce the
formalism and study the solutions and discrete symmetries at the classical
level. The constraints and corresponding dynamics are analysed in section
III. The canonical quantization of the free theory is discussed in section
IV. The chiral structure and naively renormalizable interactions are described in Section V.
We give our conclusions in section VI and close with two appendices with
technical details of the calculations.


\section{Parity-based formalism for the $(1,0) \oplus(0,1)$ representation.}

It was shown in \cite{Gomez-Avila:2013qaa} that the parity-based covariant
basis for a general $(j,0)\oplus(0,j)$ operator space contains:

\begin{enumerate}
\item Two Lorentz scalar operators, the unit matrix of dimension $2(2j+1)$
and the chirality operator $\chi$.

\item Six operators transforming in the $(1,0)\oplus(0,1)$ representation
forming a rank-2 anti-symmetric tensor, $M\indices{_\mu_\nu}$, whose
components are the corresponding generators of the HLG.

\item A pair of symmetric traceless matrix tensors transforming in the $%
(j,j)$ representation, with the first one denoted $S 
\indices{^{\mu_{1}} ^ {\mu_{2}}
^{\ldots} ^{\mu_{2j}}}$ and the second one given by $\chi S %
\indices{^{\mu_{1}} ^ {\mu_{2}} ^{\ldots} ^{\mu_{2j}}}$.

\item A series of tensor matrix operators with the appropriate symmetry
properties such that they transform in the $(2,0)\oplus(0,2), (3,0)\oplus
(0,3), ...,(2j,0)\oplus(0,2j) $ representations of the HLG.
\end{enumerate}

The rest frame parity operator is the time component of the first symmetric
traceless tensor, $\Pi=S^{00...0}$. The boost operator can be explicitly
constructed due to the simple representation form (in the chiral basis for
the $(j,0)\oplus(0,j)$ space) of the boost generator $\mathbb{K} = -i \chi 
\mathbb{J} = -i \mathrm{diag} (\mathbf{J}, -\mathbf{J})$. Using the boost
operator, it is possible to construct explicitly the states (\textit{$j$-spinors} or simply spinors in the following ) in an arbitrary frame once we
know them in the rest frame. Another important application of the boost
operator is the construction of the covariant form of a given operator from
its form in the rest frame. In particular, we can calculate the covariant
form of the parity operator. A simple calculation yields 
\begin{equation}
B(p)\Pi B^{-1}(p)=\frac{S^{\mu_{1}\mu_{2}...\mu_{2j}}p_{\mu_{1}}p_{\mu_{2}
}...p_{\mu_{2j}}}{m^{2j}}.  \label{parcov}
\end{equation}

Let us briefly review the application to $j=1/2$. In this case the covariant
basis is given by two scalar operators, $\mathbf{1}$ and $\chi$, an
antisymmetric tensor, $M\indices{_\mu_\nu}$, and and two vector operators
(the ``symmetric'' operators of rank $2j=1$). 
\begin{equation}
\{\mathbf{1},\chi,S^{\mu},\chi S^{\mu},M\indices{^\mu^\nu}\}.
\label{basis12}
\end{equation}
The algorithm outlined in \cite{Gomez-Avila:2013qaa} yields 
\begin{equation}
S^{\mu}=\Pi\left( g^{0\mu}-2iM^{0\mu}\right) .
\end{equation}
This is the conventional set used in the literature up to a $1/2$ factor in $%
M\indices{_\mu_\nu}$, where the chirality operator is the conventional $%
\gamma^{5}$ Dirac matrix and $S^{\mu}=\gamma^{\mu}$. Boosting the rest frame
parity operator we get 
\begin{equation}
B(p)\Pi B^{-1}(p)=\frac{S^{\mu}p_{\mu}}{m}.  \label{piboost}
\end{equation}
Since the rest frame projectors onto states of well-defined parity are 
\begin{equation}
\tilde{\mathbb{P}}_{\pm}=\frac{1}{2}\left( 1\pm\Pi\right),  \label{parproj}
\end{equation}
the condition for well-defined parity in the rest frame is 
\begin{equation}  \label{rfproj}
\tilde{\mathbb{P}}_{\pm}u(0)=u(0),
\end{equation}
and boosting this equation we get the following condition 
\begin{equation}
\left( S^{\mu}p_{\mu}\mp m\right) u(p)=0.
\end{equation}
Transforming to configuration space the positive parity projection yields
the Dirac equation 
\begin{equation}
\left( iS^{\mu}\partial_{\mu}- m\right) \psi(x)=0,
\end{equation}
where $\psi(x)=u(p)e^{-ip\cdot x}$.

\subsection{The structure of the spin-one representation}

In the case of spin-one, the basis of matrices with well-defined Lorentz
transformation properties is%
\begin{equation}
\{\mathbf{1},\chi,S\indices{^\mu^\nu},\chi S\indices{^\mu^\nu},M %
\indices{^\mu^\nu}, C^{\mu\nu\alpha\beta}\}.  \label{basis1}
\end{equation}
The symmetric tensor $S\indices{^\mu^\nu}$ is given by%
\begin{equation}
S\indices{^\mu^\nu}=\Pi\left( g\indices{^\mu^\nu}-i(g^{0\mu}M^{0\nu}+g^{0\nu
}M^{0\mu})-\{M^{0\mu},M^{0\nu}\}\right) .  \label{st1}
\end{equation}
This tensor is traceless in the Lorentz indices%
\begin{equation}
S\indices{^\mu_\mu}=0,
\end{equation}
which leaves nine independent components transforming in the $(1,1)$
representation of the HLG. These operators satisfy the following algebraic
relations 
\begin{align}
\lbrack S\indices{^\mu^\nu}, S\indices{^\alpha^\beta}] & = -i\left( g%
\indices{^\mu^\alpha} M\indices{^\nu^\beta} + g\indices{^\nu^\alpha} M%
\indices{^\mu^\beta} + g\indices{^\nu^\beta} M\indices{^\mu^\alpha} + g%
\indices{^\mu^\beta} M\indices{^\nu^\alpha} \right) ,  \label{CRS} \\
\left\{ S\indices{^\mu^\nu},S\indices{^\alpha^\beta}\right\} & = \frac{4}{3}%
\left( g^{\mu\alpha} g^{\nu\beta} + g^{\nu\alpha} g^{\mu\beta} -\frac{1}{2}
g^{\mu\nu} g\indices{^\alpha^\beta} \right) - \frac{1}{6} \left( C%
\indices{^\mu ^\alpha^\nu^\beta} + C\indices{^\mu^\beta^\nu^\alpha}\right) .
\label{ACRS}
\end{align}
Finally the tensor transforming in the $(2,0)\oplus(0,2)$ representation is
given by 
\begin{equation}
C^{\mu\nu\alpha\beta} =4\{M\indices{^\mu^\nu},M\indices{^\alpha^\beta}
\}+2\{M^{\mu\alpha}
,M^{\nu\beta}\}-2\{M^{\mu\beta},M^{\nu\alpha}\}-8(g^{\mu\alpha}g^{\nu\beta
}-g^{\mu\beta}g^{\nu\alpha}).  \label{Weyl}
\end{equation}
It has the following symmetries 
\begin{equation}
C_{\mu\nu\alpha\beta}=-C_{\nu\mu\alpha\beta}=-C_{\mu\nu\beta\alpha},\quad
C_{\mu\nu\alpha\beta}=C_{\alpha\beta\mu\nu},  \label{Weylsym}
\end{equation}
the contraction of any pair of indices vanishes and it satisfies the
algebraic Bianchi identity%
\begin{equation}
C_{\mu\nu\alpha\beta}+C_{\mu\alpha\beta\nu}+C_{\mu\beta\nu\alpha}=0.
\label{Bianchi}
\end{equation}
These symmetries leave only $10$ independent components out of the $256$
components of a general four-index tensor.

The explicit form of the $6\times6$ matrix tensor operators in Eq. (\ref%
{basis1}) can be found in \cite{Gomez-Avila:2013qaa}, in the chiral basis of
states diagonalizing the chirality operator, $\chi$. For the purposes of
this work it is convenient to work in the ``parity'' basis of states where
the particle-anti-particle interpretation is easier. The matrix operators
are related by $\mathcal{O}=F\mathcal{O}_{\chi}F^{\dagger}$ where $F$ stands
for the change of basis matrix 
\begin{equation}
F=\frac{1}{\sqrt{2}}\left( 
\begin{array}{cc}
I & I \\ 
I & -I%
\end{array}
\right) .
\end{equation}
Here we will just need the explicit representation of $S\indices{^\mu^\nu}$,
which in the parity basis is given by 
\begin{equation}
S^{00}\equiv\Pi=\left( 
\begin{array}{cc}
I & 0 \\ 
0 & -I%
\end{array}
\right) ,\quad S^{0i}=\left( 
\begin{array}{cc}
0 & -J^{i} \\ 
J^{i} & 0%
\end{array}
\right) ,\quad S^{ij}=\left( 
\begin{array}{cc}
g^{ij}+\left\{ J^{i},J^{j}\right\} & 0 \\ 
0 & -g^{ij}-\left\{ J^{i},J^{j}\right\}%
\end{array}
\right) ,
\end{equation}
where $J^{i}\equiv\frac{1}{2}\epsilon^{ijk}M_{jk}$ are the conventional spin
one matrices.

\subsection{The spin-one parity projection}

The condition for a state transforming in $(1,0)\oplus(0,1)$ to have
well-defined parity is given by Eq.(\ref{rfproj}), with the corresponding
parity operator in this representation space. A similar procedure as the one
used for the spin $1/2$ case yields the following equation 
\begin{equation}
\left( S^{\mu\nu}\partial_{\mu}\partial_{\nu}+m^{2}\right) \psi\left(
x\right) =0.  \label{eomw}
\end{equation}
This equation was proposed long ago by Weinberg \cite{Weinberg:1964cn}
following a different approach and several aspects of this theory have
been studied in the literature
\cite{Eeg:1972lt},\cite{Eeg:1972mt},\cite{Eeg:1973}%
, \cite{Ahluwalia:1999ny}. The main drawback of this equation is that
it contains unphysical solutions. In the parity-based covariant
construction this is easily understood from the algebraic properties
of the symmetric tensor in Eq. (\ref{ACRS}). Indeed, using this
equation it is easy to show that
\begin{equation}
(S^{\mu\nu}\partial_{\mu}\partial_{\nu})^{2}\equiv(S(\partial))^{2}
=\partial^{4},  \label{S(p)2}
\end{equation}
and multipliying on the left Eq. (\ref{eomw}) with $S(\partial)-m^{2}$ we
obtain 
\begin{equation}
\left( \partial^{4}-m^{4}\right) \psi\left( x\right) =0.  \label{d4}
\end{equation}
This equation has the conventional plane wave solutions with $p^{2}=m^{2}$
but also solutions belonging to the $p^{2}=-m^{2}$ Poincar\'e orbit. This
problem can be traced back to the naive construction of the projectors in
Eq.(\ref{parproj}). It can be shown that the corresponding boosted operators 
\begin{equation}
\tilde{\mathbb{P}}_{\pm}(\mathbf{p})=\frac{1}{2}\left( 1\pm\frac{S(p)}{m^{2}}
\right)
\end{equation}
cease to be projectors as soon as we go off-shell. The correct parity
projectors for the general off-shell case are 
\begin{equation}
\mathbb{P}_{\pm}(\mathbf{p})=\frac{1}{2}\left( 1\pm\frac{S(p)}{p^{2}}
\right) .
\end{equation}
In addition to finding the right parity projection we must also take care of
the projection on the desired Poincar\'e orbit. To this end we use the
simultaneous mass and parity projector 
\begin{equation}
\frac{p^{2}}{m^{2}}\mathbb{P}_{\pm}(\mathbf{p})=\frac{1}{2m^{2}}\left(
p^{2}\pm S(p)\right) .
\end{equation}
This procedure yields the following equation in coordinate space: 
\begin{equation}
\left( \Sigma^{\mu\nu}\partial_{\mu}\partial_{\nu}+m^{2}\right) \psi\left(
x\right) =0,  \label{eomgood}
\end{equation}
where 
\begin{equation}
\Sigma^{\mu\nu} = \frac{1}{2}\left( g^{\mu\nu}+S^{\mu\nu} \right) .
\end{equation}
Using Eq.(\ref{S(p)2}) and multiplying Eq. (\ref{eomgood}) on the left by $%
\frac{1}{2}(\partial^{2}-S(\partial) )-m^{2}$ it is easy to show that the
field satisfy the Klein-Gordon equation 
\begin{equation}
( \partial^{2} + m^{2} )\psi(x)=0,
\end{equation}
whose solutions are of the form $\psi(x)=u_{r}(\mathbf{p})e^{-ip\cdot x}$
where $r$ denotes the particle polarization. The theory for particles with
negative parity can be constructed in a similar way; in the following we
will focus on the positive parity case.

The formulation of wave equations for spinning particles is an old
problem and as far as we know Eq. (\ref{eomgood}) was firstly
considered in \cite%
{Shay:1968iq} following a different approach, including
electromagnetic interactions at the classical level. Closely related
work was also done in \cite{Hammer:1968zz}, \cite{Tucker:1971bi}. The
present approach, based on the parity and Poincar\'e projections,
permits us to identify all quantum numbers from first
principles. Also, the algebraic structure of the $%
(1,0)\oplus(0,1)$
representation space will allow us to work out the constrained
dynamics at the classical level and the proper quantization of this
theory.

The spinors $u_{r}(\mathbf{p})$ have six components and satisfy the
following equation 
\begin{equation}
\left( \Sigma^{\mu\nu}p_{\mu}p_{\nu}-m^{2}\right) u_{r}(\mathbf{\mathbf{p} }
)=0.  \label{eomugood}
\end{equation}
Equivalently, since a free particle spinor must satisfy the
Klein-Gordon condition, the spinor also satisfies
\begin{equation}
\left( S^{\mu\nu}p_{\mu}p_{\nu}-m^{2}\right) u_{r}(\mathbf{\mathbf{p}})=0.
\label{eomuw}
\end{equation}

Let us first explore the free particle solutions of Eq. (\ref{eomgood}).
Introducing the explicit form of the $S^{\mu\nu}$ matrices in Eq. (\ref%
{eomgood}) we get 
\begin{equation}
\left( 
\begin{array}{cc}
\partial^{2}+m^{2}+(\mathbf{J}\cdot\nabla)^{2} & -\mathbf{J}\cdot
\nabla\partial_{0} \\ 
\mathbf{J}\cdot\nabla\partial_{0} & m^{2}-(\mathbf{J}\cdot\nabla)^{2}%
\end{array}
\right) \psi(x)=0.
\end{equation}
Writing $\psi$ in terms of the ``up'' ($\varphi$) and ``down'' ($\xi$) three-component
components we get 
\begin{align}  \label{eq:dynsig}
\left[\partial^2 + m^2 + (\vec{J}\cdot\vec{\nabla})^2 \right] \varphi &= 
\vec{J}\cdot\vec{\nabla}\ \partial_0 \xi, \\
\left[m^2 - (\vec{J}\cdot\vec{\nabla})^2\right] \xi &= - \vec{J}\cdot\vec{%
\nabla}\ \partial_0 \varphi.
\end{align}
The second line yields the $\xi$ field in terms of the time derivatives of
the $\varphi$ field, i.e. it is a constraint of the theory which leaves only
the three complex components of $\varphi$ required to describe a
particle-antiparticle spin-one system as the physical degrees of freedom.
The constraint equation reads 
\begin{equation}  \label{eq:const}
\xi = - \mathcal{O}^{-1} \vec{J}\cdot\vec{\nabla} \partial_0 \varphi,
\end{equation}
with $\mathcal{O} = m^2 - (\vec{J}\cdot\vec{\nabla})^2$ which is
non-singular.

The true equation of motion for the $\varphi$ field is obtained multiplying
the first equation by $\mathcal{O}$ and using the second one to get 
\begin{equation}  \label{eq:trueeq}
\left( \left[\partial^2 + m^2 + (\vec{J}\cdot\vec{\nabla})^2 \right] \left[%
m^2 - (\vec{J}\cdot\vec{\nabla})^2\right] + (\vec{J}\cdot\vec{\nabla})^2\
\partial_0^2 \right) \varphi = 0.
\end{equation}

Notice that this equation is second order in time derivatives and seemingly
higher order in space derivatives. However, because of the algebraic
properties of $J_i$ matrices, 
\begin{equation}  \label{eq:jp3}
(\vec{J}\cdot\vec{\nabla})^3 = (\vec{J}\cdot\vec{\nabla}) \vec{\nabla}^2,
\end{equation}
and it is easy to show that this equation can be rewritten as 
\begin{equation}  \label{eq:trve}
m^2 \left[\partial^2 + m^2 \right] \varphi = 0,
\end{equation}
i.e., it is just the Klein-Gordon equation for the three complex degrees of
freedom in $\varphi$.

In momentum space, writing $\varphi (x)=\phi_{r}(\mathbf{p})e^{-ip\cdot x}$ we find
the following solutions to the equation of motion 
\begin{equation}
u_{r}(p)= N 
\begin{pmatrix}
{\phi_{r}(\mathbf{p})} \\ 
{-\frac{\mathbf{J}\cdot\mathbf{p}}{E}\,\phi_{r}(\mathbf{p})}%
\end{pmatrix},  
\label{spinor}
\end{equation}
where $N$ is an appropriate normalization factor.

Our formalism is designed for massive particles. However, it has a soft $m\to
0$ limit which is worth exploring. In the massless limit, our equation
reduces to the system 
\begin{align}  \label{eq:dynsig}
\left[\partial^2 + (\vec{J}\cdot\vec{\nabla})^2 \right] \varphi - \vec{J}%
\cdot\vec{\nabla}\ \partial_0 \xi &= 0   \\
\vec{J}\cdot\vec{\nabla}\ \partial_0 \varphi - (\vec{J}\cdot\vec{\nabla}%
)^2\xi &= 0.
\end{align}
Notice that now the operator $(\vec{J}\cdot\vec{\nabla})^2$
accompanying the $\xi$ spinor is not invertible (in momentum space, it
is the helicity operator, and it has a zero eigenvalue). In this case
we expect to have a gauge invariance which reduces the degrees of
freedon contained in the $\psi$ spinor. In the next section, we will
work out the Hamiltonian analysis of the constrained dynamics of the
theory, and will show that in the massive case all constraints are
second class. In the massless limit the characteristic matrix of the
constraints has no inverse and first class constraints (gauge
symmetries) appear. A straightforward calculation shows that the
massless equation of motion (or the Lagrangian in the following
section) is invariant under the following gauge transformations
\footnote{%
  The easiest way to check this is to use the Cartesian
  representations of the spin-one matrices
  $(J_i)_{jk}= -i \epsilon_{ijk}$, which yields $(\vec{J}%
  \cdot\vec{\nabla})_{ij} \partial_{j} f =0$.}
\begin{align}
\varphi_{i} &\rightarrow \varphi_{i} +  (\vec{J}\cdot\vec{\nabla})_{ij}  \varepsilon_{j} ,  \\
\xi_{i} &\rightarrow \xi_{i} + \partial^0 \varepsilon_{i} + \partial_{i} f,  
\label{gaugeinv}
\end{align}
where $\varepsilon (x)$ is an arbitrary three component spinor, and $f (x)$ is an
arbitrary scalar function. This reduces our six degrees of freedom to only
two as expected.

Coming back to the massive theory which is the topic of this paper, the
presence of non dynamical degrees of freedom in $\psi$ makes clear that the 
quantization of the theory must proceed through a careful study of the constraints. Before
elaborating on this point and in preparation for the particle interpretation
necessary for the quantization of the theory, we study the charge
conjugation operation.


\subsection{Interacting theory and discrete symmetries.}

We use the gauge principle for the simplest case of a $U(1)$ gauge group.
Gauging Eq. (\ref{eomgood}) we get 
\begin{equation}
\left[ \Sigma ^{\mu \nu }\left( i\partial -qA\right) _{\mu }\left( i\partial
-qA\right) _{\nu }-m^{2}\right] \psi =0,  \label{eomg}
\end{equation}%
where $q$ is the $U(1)$ charge of the particle. Complex conjugating Eq. (\ref%
{eomg}) and multiplying on the left by a matrix in the $(1,0)\oplus (0,1)$
representation space denoted by $\Gamma $ we obtain%
\begin{equation}
\left[ \Gamma (\Sigma ^{\mu \nu })^{\ast }\Gamma ^{-1}\left( i\partial
+qA\right) _{\mu }\left( i\partial +qA\right) _{\nu }-m^{2}\right] \psi
^{c}=0,  \label{cceom}
\end{equation}%
with 
\begin{equation}
\psi ^{c}\equiv \mathcal{C}\psi =\Gamma \psi ^{\ast }.
\end{equation}%
If we require $\psi ^{c}$ to satisfy the same equation as $\psi $ but with
the opposite $U(1)$ charge, $-q$, the symmetric tensor $S$ must satisfy the
following relation 
\begin{equation}
\Gamma (S^{\mu \nu })^{\ast }\Gamma ^{-1}=S^{\mu \nu }.  \label{stransf}
\end{equation}

The construction of the matrix $\Gamma$ satisfying Eq. (\ref{stransf}) can
be done from first principles and we just quote the final result. Up to a
phase this matrix is given by 
\begin{equation}
\Gamma=\left( 
\begin{array}{cc}
U & 0 \\ 
0 & -U%
\end{array}
\right),
\end{equation}
where $U$ stands for the time reversal operator in the $(1,0)\oplus(0,1)$
representation space: 
\begin{equation}
U=e^{-i\pi J_{2}}=\left( 
\begin{array}{ccc}
0 & 0 & 1 \\ 
0 & -1 & 0 \\ 
1 & 0 & 0%
\end{array}
\right) .
\end{equation}
A crucial difference with the Dirac theory is that for spin-one matter
fields the charge conjugation operator commutes with the rest frame parity
operator, 
\begin{equation}
\lbrack\mathcal{C},\Pi]=0.
\end{equation}
This relation defines the particle-antiparticle structure in the
corresponding quantum field theory. In the rest frame, the ``down''
component of the spinors in Eq. (\ref{spinor}) corresponds to negative
parity as in the Dirac case. However, for spin-one matter particles, it is
not connected with the antiparticle solutions. Indeed, as we can see from
the explicit form of the spinors in Eq. (\ref{spinor}), the ``down''
component vanishes in the rest frame, and for an arbitrary frame it is fixed
by the kinematics.

The charge conjugated spinor, given by 
\begin{equation}
u_{r}^{c}(\mathbf{\mathbf{p}})=\Gamma u_{r}^{\ast}(\mathbf{\mathbf{p}}),
\end{equation}
also satisfies the equation 
\begin{equation}
\left( \Sigma^{\mu\nu}p_{\mu}p_{\nu}-m^{2}\right) u_{r}^{c}(\mathbf{p})=0.
\label{eomuc}
\end{equation}
The adjoint spinors obey the adjoint equations 
\begin{align}
\bar{u}_{r} (\mathbf{p}) (S^{\mu\alpha} p_{\mu}p_{\alpha} - m^{2}) & =0,
\label{eomubar} \\
\bar{u}_{r}^{c} (\mathbf{p}) (S^{\mu\alpha} p_{\mu}p_{\alpha} - m^{2}) & =0.
\label{eomucbar}
\end{align}
These spinors are normalized according to 
\begin{equation}
\bar{u}_{r}^{c}(\mathbf{p})u_{s}^{c}(\mathbf{p})=\bar{u}_{r}(\mathbf{p}
)u_{s}(\mathbf{p})=\delta_{rs}.
\end{equation}
The corresponding completeness relation is 
\begin{equation}
{\displaystyle{\displaystyle\sum_{r}}}u_{ra}(\mathbf{p})\bar{u}_{rb} ( 
\mathbf{p})={\displaystyle{\displaystyle\sum_{r}}}u_{ra}^{c}(\mathbf{p} ) 
\bar{u}_{rb}^{c}(\mathbf{p})=\left( \frac{S\left( \mathbf{p}\right) +m^{2}}{
2m^{2}}\right) _{ab}.  \label{eq:espinoresproyector}
\end{equation}

Now, the minimally coupled equation, Eq. (\ref{eomg}), written in terms of the
covariant derivative 
\begin{equation}  \label{eq:minicoupl}
D_{\mu }\psi = \partial _{\mu }\psi + i q A_{\mu }\psi ,
\end{equation}
and the parity components $\{\varphi, \xi\}$, is 
\begin{align}
\left( D^2 + m^2 + \frac{1}{2} D_i\{J_i,J_j\}D_j \right) \varphi - \frac{1}{2%
} J_i\{D_i,D_0\}\xi &= 0,  \label{eq:eomgD} \\
\frac{1}{2} J_i\{D_i,D_0\}\varphi + \left( m^2 - \frac{1}{2}
D_i\{J_i,J_j\}D_j \right) \xi & = 0 .  \label{eq:gDcon}
\end{align}
Again, Eq. (\ref{eq:gDcon}) does not involve the time derivative of $%
\xi$ and is therefore still a constraint. While the manipulation of this
equation is complicated by the presence of the non-commuting differential
operators $D^\mu$, we can check in a calculation similar to the one leading
to Eq. (\ref{eq:trueeq}) that the true equation of motion has the form 
\begin{equation}  \label{eq:trueeqg}
\left[ \left( D^2 + m^2 + \frac{1}{2}D_i\{J_i,J_j\}D_j \right ) + \frac{1}{2}%
J_i\{D_i,D_0\} \mathcal{O}_c^{-1} \frac{1}{2}J_j\{D_j,D_0\} \right] \varphi
=0 .
\end{equation}
The operator $\mathcal{O}_c^{-1} = \left[m^2 - \frac{1}{2} D_i\{J_i,
J_j\}D_j \right]^{-1}$ involves only the spatial components of $D^\mu$, and
therefore this is an equation containing only second time derivatives of the 
$\varphi$ components. Therefore, the counting of degrees of freedom is
unaltered from the free case.


\section{Classical field theory and constraints}


The equation of motion can be derived from the following Hermitian
Lagrangian 
\begin{equation}
\mathcal{L}=\partial _{\mu }\bar{\Psi}\Sigma ^{\mu \nu }\partial _{\nu }\Psi
-m^{2}\bar{\Psi}\Psi  \label{freelag}
\end{equation}%
where $\bar{\Psi}=\Psi ^{\dagger }\Pi $. In order to exhibit the dynamical
content we write the theory in terms of the \textquotedblleft
up\textquotedblright\ ($\varphi $) and \textquotedblleft
down\textquotedblright\ ($\xi $) components of the field and the
corresponding conjugate momenta 
\begin{equation}
\Psi =\left( 
\begin{array}{c}
\varphi \\ 
\xi%
\end{array}
\right) ,\qquad \zeta =\left( 
\begin{array}{c}
\pi \\ 
\tau%
\end{array}
\right) .
\end{equation}%
In terms of these components the Lagrangian reads 
\begin{align}
\mathcal{L}& =\partial _{0}\varphi ^{\dagger }\partial _{0}\varphi +\partial
_{i}\varphi ^{\dagger }\partial ^{i}\varphi -\frac{1}{2}\partial _{0}\varphi
^{\dagger }J^{i}\partial _{i}\xi -\frac{1}{2}\partial _{0}\xi ^{\dagger
}J^{i}\partial _{i}\varphi -\frac{1}{2}\partial _{i}\varphi ^{\dagger
}J^{i}\partial _{0}\xi -\frac{1}{2}\partial _{i}\xi ^{\dagger }J^{i}\partial
_{0}\varphi +\frac{1}{2}\partial _{i}\varphi ^{\dagger }\left\{
J^{i},J^{j}\right\} \partial _{j}\varphi  \notag \\
& +\frac{1}{2}\partial _{i}\xi ^{\dagger }\left\{ J^{i},J^{j}\right\}
\partial _{j}\xi -m^{2}\left( \varphi ^{\dagger }\varphi -\xi ^{\dagger }\xi
\right) .  \label{DLagrangianaUD}
\end{align}%
Notice that this lagrangian does not contain second time derivatives in the
\textquotedblleft down\textquotedblright\ component $\xi $. The canonical
conjugated momenta are given as 
\begin{align}
\pi _{a}& =\frac{\delta \mathcal{L}}{\delta \left( \partial _{0}\varphi
_{a}\right) }=\partial _{0}\varphi _{a}^{\dagger }-\frac{1}{2}\left(
\partial _{i}\xi ^{\dagger }J^{i}\right) _{a},\quad \pi _{a}^{\dagger }=%
\frac{\delta \mathcal{L}}{\delta \left( \partial _{0}\varphi _{a}^{\dagger
}\right) } =\partial _{0}\varphi _{a}-\frac{1}{2}\left( J^{i}\partial
_{i}\xi \right) _{a},  \label{Momento_psi} \\
\tau _{a}& =\frac{\delta \mathcal{L}}{\delta \left( \partial _{0}\xi
_{a}\right) }=-\frac{1}{2}\left( \partial _{i}\varphi ^{\dagger
}J^{i}\right) _{a},\qquad \qquad \tau _{a}^{\dagger }=\frac{\delta \mathcal{L%
}}{\delta \left( \partial _{0}\xi _{a}^{\dagger }\right) }=-\frac{1}{2}%
\left( J^{i}\partial _{i}\varphi \right) _{a}.  \label{Momento_tau}
\end{align}%
Clearly, Eqs. (\ref{Momento_tau}) are (primary) constraints on the variables
of the system 
\begin{equation}
\rho _{a}=\tau _{a}+\frac{1}{2}\left( \partial _{i}\varphi ^{\dagger
}J^{i}\right) _{a}=0,\qquad \rho _{a}^{\dagger }=\tau _{a}^{\dagger }+\frac{%
1 }{2}\left( J^{i}\partial _{i}\varphi \right) _{a}=0.
\label{rhoconstraints}
\end{equation}

The hamiltonian density is 
\begin{equation}
\mathcal{H}\mathcal{=}\pi_{a}\partial_{0}\varphi_{a}+\partial_{0}\varphi
_{a}^{\dagger}\pi_{a}^{\dagger}+\tau_{a}\partial_{0}\xi_{a}+\partial_{0}
\xi_{a}^{\dagger}\tau_{a}^{\dagger} - \mathcal{L}.  \label{Hdensity}
\end{equation}
A straightforward calculation yields 
\begin{align}
\mathcal{H} & \mathcal{=}\pi_{a}\pi_{a}^{\dagger}+\frac{1}{2}\pi_{a}\left(
J^{i}\partial_{i}\xi\right) _{a}+\frac{1}{2}\left( \partial_{i}\xi^{\dagger
}J^{i}\right) _{a}\pi_{a}^{\dagger}+\frac{1}{4}\left( \partial_{i}
\xi^{\dagger}J^{i}\right) _{a}\left( J^{j}\partial_{j}\xi\right) _{a}  \notag
\\
& -\partial_{i}\varphi_{a}^{\dagger}\partial^{i}\varphi_{a}-\frac{1}{2}%
\partial _{i}\varphi_{a}^{\dagger}\left\{ J^{i},J^{j}\right\}
_{ab}\partial_{j}\varphi _{b}- \frac{1}{2}\partial_{i}\xi_{a}^{\dagger}\left%
\{ J^{i},J^{j}\right\} _{ab}\partial_{j}\xi_{a}+m^{2}\left(
\varphi_{a}^{\dagger}\varphi_{a}-\xi _{a}^{\dagger}\xi_{a}\right) .
\end{align}
Notice that this Hamiltonian density does not contain the $\tau$ momenta nor
time derivatives of the ``down'' spinor. According to Dirac classic lectures 
\cite{Dirac:1964} the time evolution of the system is given by the modified
Hamiltonian $\mathcal{H}^{\ast}$ given by 
\begin{equation}
H^{\ast}=\int d^{3}x \mathcal{H}^{\ast}.  \label{Hamiltoniano_Mofificado}
\end{equation}
with the modified Hamiltonian density 
\begin{equation}
\mathcal{H}^{\ast} =\mathcal{H}+\lambda_{a}\rho_{a} + \lambda_{a}^{\dagger
}\rho_{a}^{\dagger},  \label{Dhamiltoniana_Modificada}
\end{equation}
where $\lambda_{a}$ and $\lambda_{a}^{\dagger}$ are the Lagrange multipliers.

The Hamilton equations read 
\begin{align}
\partial_{0}\varphi_{a} & =\frac{\delta H^{\ast}}{\delta\pi_{a}}=\pi
_{a}^{\dagger}+\frac{1}{2}\left( J^{i}\partial_{i}\xi\right) _{a}, \\
\partial_{0}\pi_{a} & =-\frac{\delta H^{\ast}}{\delta\varphi_{a}}=-\partial
_{i}\partial^{i}\varphi_{a}^{\dagger}-\partial_{j}\partial_{i}\left(
\varphi^{\dagger}J^{i}J^{j}\right) _{a}-m^{2}\varphi_{a}^{\dagger}+\frac{1} {%
2} \left( \partial_{i}\lambda^{\dagger}J^{i}\right) _{a}, \\
\partial_{0}\xi_{a} & =\frac{\delta H^{\ast}}{\delta\tau_{a}}=\lambda_{a}, \\
\partial_{0}\tau_{a} & =-\frac{\delta H^{\ast}}{\delta\xi_{a}}=\frac{1} {2}
\partial_{i}\left( \pi J^{i}\right) _{a}-\frac{3}{4}\left( \partial
_{j}\partial_{i}\xi^{\dagger}J^{i}J^{j}\right) _{a}+m^{2}\xi_{a}^{\dagger}.
\end{align}
The corresponding equations for the adjoint phase space variables, not shown
here, are given by the adjoint of these equations.

The time evolution of any observable can be written in terms of the Poisson
brackets as 
\begin{equation}
\dot{A}=\left\{ A,H^{\ast }\right\} .
\end{equation}

In our case the Poisson bracket is given by 
\begin{equation}
\left\{ A\left( \mathbf{x}\right) ,B\left( \mathbf{y}\right) \right\} =\int
d^{3}\mathbf{x}^{\prime}\sum_{a}\left[ \frac{\delta A\left( \mathbf{x}
\right) }{\delta\Psi_{a}\left( \mathbf{x}^{\prime}\right) } \frac{\delta
B\left( \mathbf{y}\right) }{\delta\zeta_{a} \left( \mathbf{x}^{\prime}
\right) }-\frac{\delta B\left( \mathbf{y}\right) }{\delta\Psi_{a}\left( 
\mathbf{x}^{\prime}\right) }\frac{\delta A\left( \mathbf{x} \right) }{
\delta\zeta_{a}\left( \mathbf{x}^{\prime}\right) }\right]
\label{Patentesis_Poisson}
\end{equation}
where the sum is over all the field components and their conjugate momenta.

A straightforward calculation yields 
\begin{equation}
\left\{ \varphi _{a}\left( \mathbf{x}\right) ,\pi _{b}\left( \mathbf{y}%
\right) \right\} =\delta _{ab}\delta ^{3}\left( \mathbf{x-y}\right) ,\qquad
\left\{ \xi _{a}\left( \mathbf{x}\right) ,\tau _{b}\left( \mathbf{y}\right)
\right\} =\delta _{ab}\delta ^{3}\left( \mathbf{x-y}\right) ,
\label{Poisson_B_Fields}
\end{equation}%
and the corresponding adjoint relations.

The dynamics generated by $H^{\ast }$ must preserve the constraints hence
the following relations must hold 
\begin{equation}
\partial _{0}\rho _{a}=\left\{ \rho _{a},H^{\ast }\right\} =0,\qquad
\partial _{0}\rho _{a}^{\dagger }=\left\{ \rho _{c}^{\dagger },H^{\ast
}\right\} =0.
\end{equation}%
In our system this produces new (secondary) constraints 
\begin{align}
\kappa _{a}& =\partial _{i}\left( \pi J^{i}\right) _{a}-\frac{1}{2}\left(
\partial _{j}\partial _{i}\xi ^{\dagger }J^{i}J^{j}\right) _{a}+m^{2}\xi
_{a}^{\dagger }=0, \\
\kappa _{a}^{\dagger }& =\partial _{i}\left( J^{i}\pi ^{\dagger }\right)
_{a}-\frac{1}{2}\left( \partial _{j}\partial _{i}J^{i}J^{j}\xi \right)
_{a}+m^{2}\xi _{a}=0.
\end{align}%
Requiring that the new constraints be preserved by the dynamics we get 
\begin{equation}
\lambda _{a}^{\dagger }-\partial _{i}\left( \varphi ^{\dagger }J^{i}\right)
_{a}=0,\qquad \lambda _{a}-\partial _{i}\left( J^{i}\varphi \right) _{a}=0.
\end{equation}%
These relations just define the Lagrange multipliers but do not generate new
constraints.

In total, we have twenty-four degrees of freedom in the Hamiltonian
description, twelve coming from the $\{\varphi^\dagger,\varphi\}$ fields and
their associated momenta $\{\pi^\dagger, \pi\}$, and another twelve from the 
$\{\xi^\dagger, \xi\}$ fields and their momenta $\{\tau^\dagger, \tau\}$. On
the other hand, we have the set of twelve constraints $\left\{ f_{a}\right\}
= \{ \rho_{a} ,\rho^{\dagger}_{a}, \chi_{a}, \chi^{\dagger}_{a} \}$. This
leaves us with twelve degrees of freedom in phase space, that correspond to
three complex degrees of freedom obeying a second-degree equation of motion,
as expected for a particle-antiparticle field with three degrees of freedom.

Following the procedure outlined by Dirac in Ref. (\cite{Dirac:1964}) we
calculate now the matrix of the Poisson brackets of the constraints 
\begin{equation}
\Delta_{ab}\left( \mathbf{x},\mathbf{y}\right) =\{f_{a}(\mathbf{x} ),f_{b}( 
\mathbf{y}) \}.
\end{equation}
A straightforward calculation yields the following block matrix form 
\begin{equation}
\Delta\left( \mathbf{x},\mathbf{y}\right) = m^{2} \delta^{3}\left( \mathbf{\
x-y}\right) \left( 
\begin{array}{cccc}
0 & 0 & 0 & -\mathbf{1} \\ 
0 & 0 & -\mathbf{1} & 0 \\ 
0 & \mathbf{1} & 0 & 0 \\ 
\mathbf{1} & 0 & 0 & 0%
\end{array}
\right) .  \label{Delta}
\end{equation}
This is a non-singular matrix thus all the obtained constraints are \emph{\
second class constraints}. The inverse of this matrix is given by 
\begin{equation}
\Delta^{-1}\left( \mathbf{y},\mathbf{z}\right) = \frac{1}{m^{2}}\delta
^{3}\left( \mathbf{y-z}\right) \left( 
\begin{array}{cccc}
0 & 0 & 0 & \mathbf{1} \\ 
0 & 0 & \mathbf{1} & 0 \\ 
0 & -\mathbf{1} & 0 & 0 \\ 
-\mathbf{1} & 0 & 0 & 0%
\end{array}
\right) .  \label{invDelta}
\end{equation}

To proceed with the quantization we need the Dirac bracket, defined as 
\begin{equation}
\left\{ A,B\right\} _{D}=\left\{ A,B\right\} -\int d^{3}\mathbf{z}d^{3}%
\mathbf{z}^{\prime }\left\{ A,f_{a}\left( \mathbf{z}\right) \right\} \Delta
_{ab}^{-1}(\mathbf{z},\mathbf{z}^{\prime })\left\{ f_{b}\left( \mathbf{z}%
^{\prime }\right) ,B\right\} .
\end{equation}%
For the canonical variables the inverse matrix in Eq. (\ref{invDelta})
simplifies the calculation. For example 
\begin{align*}
\left\{ \varphi _{a}(\mathbf{x}),\pi _{b}\left( \mathbf{y}\right) \right\}
_{D}& =\delta _{ab}\delta ^{3}\left( \mathbf{x-y}\right) \\
& -\frac{1}{m^{2}}\int d^{3}\mathbf{z}\left\{ \varphi _{a}(\mathbf{x}),\rho
_{c}\left( \mathbf{z}\right) \right\} \left\{ \kappa _{c}^{\dagger }\left( 
\mathbf{z}\right) ,\pi _{b}\left( \mathbf{y}\right) \right\} \\
& -\frac{1}{m^{2}}\int d^{3}\mathbf{z}\left\{ \varphi _{a}(\mathbf{x}),\rho
_{c}^{\dagger }\left( \mathbf{z}\right) \right\} \left\{ \kappa _{c}\left( 
\mathbf{z}\right) ,\pi _{b}\left( \mathbf{y}\right) \right\} \\
& +\frac{1}{m^{2}}\int d^{3}\mathbf{z}\left\{ \varphi _{a}(\mathbf{x}%
),\kappa _{c}\left( \mathbf{z}\right) \right\} \left\{ \rho _{c}^{\dagger
}\left( \mathbf{z}\right) ,\pi _{b}\left( \mathbf{y}\right) \right\} \\
& +\frac{1}{m^{2}}\int d^{3}\mathbf{z}\left\{ \varphi _{a}(\mathbf{x}%
),\kappa _{c}^{\dagger }\left( \mathbf{z}\right) \right\} \left\{ \rho
_{c}\left( \mathbf{z}\right) ,\pi _{b}\left( \mathbf{y}\right) \right\} ,
\end{align*}%
and similar expressions hold for the remaining pairs of conjugate variables.
A straightforward calculation yields 
\begin{align}
\left\{ \varphi _{a}(\mathbf{x}),\pi _{b}\left( \mathbf{y}\right) \right\}
_{D}& =\left[ 1-\frac{\left( \mathbf{J}\cdot \nabla \right) ^{2}}{2m^{2}}%
\right] _{ab}\delta ^{3}\left( \mathbf{x-y}\right) , \\
\left\{ \varphi _{a}(\mathbf{x}),\tau _{b}\left( \mathbf{y}\right) \right\}
_{D}& =0, \\
\left\{ \xi _{a}(\mathbf{x}),\pi _{b}\left( \mathbf{y}\right) \right\} _{D}&
=0, \\
\left\{ \xi _{a}(\mathbf{x}),\tau _{b}\left( \mathbf{y}\right) \right\}
_{D}& =\frac{\left( \mathbf{J}\cdot \nabla \right) _{ab}^{2}}{2m^{2}}\delta
^{3}\left( \mathbf{x}-\mathbf{y}\right) .
\end{align}%
We can rewite these relations in compact spinor notation 
\begin{equation}
\left\{ \Psi _{a}(\mathbf{x}),\zeta _{b}\left( \mathbf{y}\right) \right\}
_{D}=\left[ \Sigma ^{00}-\frac{\left( \mathbf{J}\cdot \nabla \right) ^{2}}{%
2m^{2}}S^{00}\right] _{ab}\delta ^{3}\left( \mathbf{x-y}\right) .
\label{Ccomm}
\end{equation}

The quantization of the theory must be done replacing the Dirac bracket by
the quantum commutator $-i \left[\ ,\ \right] $, and we expect the quantum
commutator of the canonical conjugate fields to be 
\begin{equation}
\left[ \Psi_{a}(\mathbf{x}),\zeta_{b}\left( \mathbf{y}\right) \right] =i %
\left[ \Sigma^{00}-\frac{\left( \mathbf{J} \cdot\nabla\right) ^{2} }{2m^{2}}
S^{00}\right] _{ab}\delta^{3}\left( \mathbf{x-y}\right) .  \label{Qcomm}
\end{equation}

To end this section we would like to remark that the coupling to an external
$U(1)$ field can spoil the quantization procedure rendering the commutation
relations of the canonical variables ill-defined for some values of the
external field \cite{Johnson:1960vt}. We do not expect this to be the case
here as pointed by the coupled true equation of motion (\ref{eq:trueeqg})
but in order to ensure this, we performed the analogous calculations for the
coupled theory finding the very same canonical commutation relations. The
calculations are rather long, and we defer the details to Appendix \ref{app:u1coupled}. 

\section{Canonical quantization of spin 1 matter fields}

Under an infinitesimal transformation 
\begin{equation}
\Psi\rightarrow\Psi^{\prime}=\Psi+\delta\Psi  \label{psivar}
\end{equation}
the lagrangian changes as%
\begin{equation}
\delta\mathcal{L}=\partial_{\mu}\left[ \partial_{\alpha}\bar{\Psi}
\Sigma^{\alpha\mu}\delta\Psi+\delta\bar{\Psi}\Sigma^{\mu\alpha}\partial
_{\alpha}\Psi\right] .  \label{varlag}
\end{equation}

Invariance under a given transformation yields conserved currents. First,
our Lagrangian is invariant under the global $U(1)$ transformations $%
\Psi^{\prime }=e^{iq\lambda}\Psi$. The corresponding conserved current is
given by 
\begin{equation}
J^{\alpha}=iq\left( (\partial_{\mu}\bar{\Psi})\Sigma^{\mu\alpha}\Psi -\bar{
\Psi}\Sigma^{\alpha\nu}(\partial_{\nu}\Psi)\right) .
\label{eq:currentspinone}
\end{equation}

Invariance under space-time translations yields the following stress tensor 
\begin{equation}
T\indices{^\mu_\nu}=\partial_{\nu}\bar{\Psi}\Sigma^{\mu\alpha}\partial
_{\alpha}\Psi+\partial_{\alpha}\bar{\Psi}\Sigma^{\alpha\mu}\partial_{\nu}
\Psi-\eta_{\nu}^{\mu}\left( \partial_{\alpha}\bar{\Psi}\Sigma^{\alpha\beta
}\partial_{\beta}\Psi-m^{2}\bar{\Psi}\Psi\right) .
\end{equation}
The angular momentum density is similarly obtained as 
\begin{equation}
\mathit{\mathcal{M}}^{0ij}=T^{0j}x^{i}-T^{0i}x^{j}+i\left( \bar{\Psi}
\epsilon_{ijk}J_{k}\Sigma^{0\nu}\partial_{\nu}\Psi\right) -i\left(
\partial_{\mu}\bar{\Psi}\Sigma^{\mu0}\epsilon_{ijk}J_{k}\Psi\right) .
\end{equation}

The field and its adjoint are expanded in the conventional Fourier series 
\begin{align}
\Psi(x) & ={\displaystyle{\displaystyle\sum_{\mathbf{p},r}}}\alpha (\mathbf{%
p })\left[ c_{r}(\mathbf{p})u_{r}(\mathbf{p})e^{-ipx}+d_{r} ^{+}(\mathbf{p}
)u_{r}^{c}(\mathbf{p})e^{ipx}\right] ,  \label{eq:fourier} \\
\bar{\Psi}(x) & ={\displaystyle{\displaystyle\sum_{\mathbf{p},r}}} \alpha( 
\mathbf{p})\left[ c_{r}^{+}(\mathbf{p})\bar{u_{r}}(\mathbf{p}
)e^{ipx}+d_{r}( \mathbf{p})\bar{u}_{r}^{c}(\mathbf{p})e^{-ipx}\right] ,
\label{eq:fourierbar}
\end{align}
where $\alpha(\mathbf{p})=1/\sqrt{2E(\mathbf{p})V}$ and $r$ denotes the
polarization of the one-particle states. The particle (antiparticle)
creation (annihilation) operators satisfy the following commutation
relations 
\begin{equation}
\begin{array}{ccccc}
\left[ c_{r}\left( \mathbf{p}\right) ,c_{s}^{\dagger}\left( \mathbf{\
p^{\prime}}\right) \right] =\delta_{rs}\delta_{\mathbf{p} \mathbf{p}
^{\prime}}, &  &  &  & \left[ d_{r}\left( \mathbf{p}\right)
,d_{s}^{\dagger}\left( \mathbf{p^{\prime}}\right) \right] =\delta
_{rs}\delta_{\mathbf{p}\mathbf{p}^{\prime}}%
\end{array}
.  \label{eq:conmutationrelations}
\end{equation}

\subsection{Commutation relations}

The conjugated momenta are given by 
\begin{align}
\bar{\zeta}_{d} & =\frac{\partial\mathcal{L}}{\partial\bar{\Psi}_{d,0}}
=\Sigma_{da}^{0\mu}\left( \partial_{\mu}\Psi\right) _{a} ,
\label{conjmombarpi} \\
\zeta_{d} & =\frac{\partial\mathcal{L}}{\partial\Psi_{d,0}}=\left(
\partial_{\mu}\bar{\Psi}\right) _{a}\Sigma_{ad}^{\mu0}.  \label{conjmompi}
\end{align}
The commutators of the fields with the canonical conjugated momenta are
given by 
\begin{align}
\left[ \zeta_{d},\Psi_{b}\right] & =\left( \partial_{\mu}\bar{\Psi}\right)
_{a}\Sigma_{ad}^{\mu0}\Psi_{b}-\Psi_{b}\left( \partial_{\mu}\bar{\Psi }
\right) _{a}\Sigma_{ad}^{\mu0},  \label{eq:commutatormomentafield} \\
\left[ \bar{\zeta}_{d},\bar{\Psi}_{b}\right] & =\Sigma_{da}^{0\mu}\left(
\partial_{\mu}\Psi\right) _{a}\bar{\Psi}_{b}-\bar{\Psi}_{b}\Sigma_{da}^{0\mu
}\left( \partial_{\mu}\Psi\right) _{a}.  \label{eq:commutatormomentafieldbar}
\end{align}
Inserting the Fourier series in Eq. (\ref{eq:commutatormomentafield}) we get 
\begin{equation}
\left[ \zeta_{d}\left( x_{1}\right) ,\Psi_{b}\left( x_{2}\right) \right] ={\ %
\displaystyle{\displaystyle\sum_{\mathbf{p},r}}}\frac{-ip_{\mu}}{2Vp_{0} } %
\left[ \bar{u}_{ra}(\mathbf{p})u_{rb}(\mathbf{p})\Sigma_{ad}^{\mu
0}e^{ip\left( x_{1}-x_{2}\right) }+\bar{u}_{ra}^{c}(\mathbf{p})u_{rb} ^{c}( 
\mathbf{p})\Sigma_{ad}^{\mu0}e^{ip\left( x_{2}-x_{1}\right) }\right] .
\label{eq:commutatormomentum}
\end{equation}
For equal time $x_{1}^{0}=x_{2}^{0}=0$, using Eq. (\ref%
{eq:espinoresproyector}) we get 
\begin{equation}
\begin{split}
\left[ \zeta_{d}\left( \mathbf{x}_{1}\right) ,\Psi_{b}\left( \mathbf{x}
_{2}\right) \right] _{x_{1,2}^{0}=0}={} & -i{\displaystyle{\displaystyle %
\sum_{\mathbf{p}}}}\frac{p_{\mu}}{2Vp_{0} }\left( \frac{S\left( \mathbf{p}
\right) +m^{2}}{2m^{2}}\right) _{ba} \Sigma_{ad}^{\mu0}e^{ip_{i}\left(
x_{1}^{i}-x_{2}^{i}\right) } \\
& -i{\displaystyle{\displaystyle\sum_{\mathbf{p}}}}\frac{p_{\mu}}{2Vp_{0} }
\left( \frac{S\left( \mathbf{p}\right) +m^{2}}{2 m^{2}}\right)
_{ba}\Sigma_{ad}^{\mu0}e^{-ip_{i}\left( x_{1}^{i}-x_{2}^{i}\right) }.
\end{split}%
\end{equation}
Changing $p_{i}\rightarrow-p_{i}$ in the second term we get 
\begin{equation}
\left[ \zeta_{d}\left( \mathbf{x}_{1}\right) ,\Psi_{b}\left( \mathbf{x}
_{2}\right) \right] _{x_{1,2}^{0}=0}=-i{\displaystyle{\displaystyle\sum _{ 
\mathbf{p}}}}\frac{e^{ip_{i}\left( x_{1}^{i}-x_{2}^{i}\right) }}{V}\left( 
\frac{\Sigma^{00}p_{0}p_{0}+\left( 2\Sigma^{0i}\Sigma^{0j}+\Sigma^{ij}
\Sigma^{00}\right) p_{i}p_{j}}{m^{2}}\right) _{bd}.
\label{eq:commutatoravancedoementum}
\end{equation}
This equation can be further reduced using the algebra satisfied by $S$.
Indeed, using Eq. (\ref{ACRS}) it is possible to show that 
\begin{equation}
\left( 2\Sigma^{0i}\Sigma^{0j}+\Sigma^{ij}\Sigma^{00}\right) p_{i} p_{j}= 
\frac{1}{2}\left( \Sigma^{ij}p_{i}p_{j}-\mathbf{p}^{2}\Sigma ^{00}\right) .
\end{equation}
Using this relation we can further reduce our commutator to 
\begin{equation}
\left[ \zeta_{d}\left( \mathbf{x}_{1}\right) ,\Psi_{b}\left( \mathbf{x}
_{2}\right) \right] _{x_{1,2}^{0}=0}=-i{\displaystyle{\displaystyle\sum _{ 
\mathbf{p}}}}\frac{e^{ip_{i}\left( x_{1}^{i}-x_{2}^{i}\right) }}{V}\left(
\Sigma^{00}+\frac{\left( S^{ij}-g^{ij}S^{00}\right) p_{i}p_{j}}{4m^{2} }
\right) _{bd} . \label{commpipsi}
\end{equation}
Finally, using the explicit representation of the $S^{\mu\nu}$ matrices it
can be shown that 
\begin{equation}
\left( S^{ij}-g^{ij}S^{00}\right) p_{i}p_{j}=2(\mathbf{J}\cdot \mathbf{p}
)^{2}S^{00},
\end{equation}
and putting it all together we obtain 
\begin{equation}
\left[ \zeta_{d}\left( \mathbf{x}_{1}\right) ,\Psi_{b}\left( \mathbf{x}
_{2}\right) \right] _{x_{1,2}^{0}=0}=-i\left( \Sigma^{00}-\frac {(\mathbf{J}
\cdot\mathbf{\nabla})^{2}}{2m^{2}}S^{00}\right) _{bd} \delta\left( \mathbf{x}
_{1}-\mathbf{x}_{2}\right) .
\end{equation}
A similar calculation yields 
\begin{equation}
\left[ \bar{\zeta}_{d}\left( \mathbf{x}_{1}\right) ,\bar{\Psi}_{b}\left( 
\mathbf{x}_{2}\right) \right] _{x_{1,2}^{0}=0}=-i\left( \Sigma^{00} -\frac{( 
\mathbf{J}\cdot\mathbf{\nabla})^{2}}{2m^{2}}S^{00}\right) _{bd} \delta\left( 
\mathbf{x}_{1}-\mathbf{x}_{2}\right) .
\end{equation}
This is exactly the result expected from our classical analysis of the
constrained dynamics in the previous section summarized in Eq. (\ref{Ccomm}).

\subsection{Energy and momentum of the field}

The energy density of the field is defined as 
\begin{equation}
\mathcal{H}=T^{00}=\partial _{0}\bar{\Psi}\Sigma ^{00}\partial _{0}\Psi
-\partial _{i}\bar{\Psi}\Sigma ^{ij}\partial _{j}\Psi +m^{2}\bar{\Psi}\Psi .
\end{equation}%
After a straightforward algebra, integrating the normal product of $T^{00}$
we get the following expression for the total energy of the field 
\begin{equation}
\begin{split}
H={}& {\displaystyle{\displaystyle(2\pi )^{3}\sum_{p,r}}}\sum_{r^{\prime
}}\alpha (\mathbf{p})^{2}\bigl[c_{r}^{+}(\mathbf{p})c_{r^{\prime }}(\mathbf{%
p })\bar{u_{r}}(\mathbf{p})\left( \Sigma ^{00}p_{0}p{}_{0}-\Sigma
^{ij}p_{i}p{}_{j}+m^{2}\right) u_{r^{\prime }}(\mathbf{p}^{\prime }) \\
& +d_{r}(\mathbf{p})c_{r^{\prime }}(\mathbf{-p})\bar{u}_{r}^{c}(\mathbf{p}
)\left( -\Sigma ^{00}p_{0}p{}_{0}-\Sigma ^{ij}p_{i}p{}_{j}+m^{2}\right)
u_{r^{\prime }}(\mathbf{-p})e^{-2ip_{0}x^{0}} \\
& +c_{r}^{+}(\mathbf{p})d_{r^{\prime }}^{+}(-\mathbf{p})\bar{u_{r}}(\mathbf{%
p })\left( -\Sigma ^{00}p_{0}p{}_{0}-\Sigma ^{ij}p_{i}p{}_{j}+m^{2}\right)
u_{r^{\prime }}^{c}(\mathbf{-p})e^{2ip_{0}x^{0}} \\
& +d_{r}(\mathbf{p})d_{r^{\prime }}^{+}(-\mathbf{p})\bar{u}_{r}^{c}(\mathbf{%
p })\left( \Sigma ^{00}p_{0}p{}_{0}-\Sigma ^{ij}p_{i}p{}_{j}+m^{2}\right)
u_{r^{\prime }}^{c}(\mathbf{-p})\bigr].
\end{split}%
\end{equation}%
Next, we use 
\begin{equation}
-\Sigma ^{00}p_{0}p_{0}-\Sigma ^{ij}p_{i}p_{j}=2\Sigma
^{0i}p_{0}p_{i}-\Sigma \left( \mathbf{p}\right) ,
\end{equation}%
and the equations of motion in Eqs. (\ref{eomugood},\ref{eomuc},\ref{eomubar}%
,\ref{eomucbar}) to obtain 
\begin{equation}
\begin{split}
H={}& {\displaystyle{\displaystyle(2\pi )^{3}\sum_{p,r}}}\sum_{r^{\prime
}}\alpha (\mathbf{p})^{2}\bigl[c_{r}^{+}(\mathbf{p})c_{r^{\prime }}(\mathbf{%
p })\bar{u_{r}}(\mathbf{p})\left( 2\Sigma ^{0\mu }p_{0}p{}_{\mu }\right)
u_{r^{\prime }}(\mathbf{p}) \\
& +d_{r}(\mathbf{p})c_{r^{\prime }}(\mathbf{-p})\bar{u}_{r}^{c}(\mathbf{p}
)\left( 2\Sigma ^{0i}p_{0}p_{i}\right) u_{r^{\prime }}(\mathbf{-p}
)e^{-2ip_{0}x^{0}} \\
& +c_{r}^{+}(\mathbf{p})d_{r^{\prime }}^{+}(-\mathbf{p})\bar{u_{r}}(\mathbf{%
p })\left( 2\Sigma ^{0i}p_{0}p_{i}\right) u_{r^{\prime }}^{c}(\mathbf{-p}
)e^{2ip_{0}x^{0}} \\
& +d_{r^{\prime }}^{+}(\mathbf{p})d_{r}(\mathbf{p})\bar{u}_{r}^{c}(\mathbf{p}
)\left( 2\Sigma ^{0\mu }p_{0}p{}_{\mu }\right) u_{r^{\prime }}^{c}(\mathbf{p}
)\bigr].
\end{split}%
\end{equation}%
With the aid of Eqs. (\ref{CRS},\ref{ACRS}) it is possible to show that 
\begin{align}
\bar{u_{r}}(\mathbf{p})\left( \Sigma ^{0\mu }p{}_{\mu }\right) u_{s}(\mathbf{%
\ p})& =p^{0}\delta _{rs},  \label{eq:productimportant} \\
\bar{u_{r}}(\mathbf{p})\left( \Sigma ^{0i}p_{i}\right) u_{r^{\prime }}^{c}( 
\mathbf{-p})& =\bar{u}_{r}^{c}(\mathbf{p})\left( \Sigma
^{0i}p_{0}p_{i}\right) u_{r^{\prime }}(\mathbf{-p})=0.
\label{eq:productimportant2}
\end{align}%
Using these results we obtain the expected total energy of the field: 
\begin{equation}
H={\displaystyle{\displaystyle\frac{(2\pi )^{3}}{V}\sum_{p,r}}p_{0}}
[c_{r}^{+}(\mathbf{p})c_{r}(\mathbf{p})+d_{r}^{+}(\mathbf{p})d_{r}(\mathbf{p}
)].
\end{equation}

The total momentum of the field is 
\begin{equation}
P_{i}=\int N\{T_{~i}^{0}\}d^{3}x=\int N\{\partial_{i}\bar{\Psi}\Sigma^{0\nu
}\partial_{\nu}\Psi+\partial_{\mu}\bar{\Psi}\Sigma^{\mu0}\partial_{i}
\Psi\}\}d^{3}x.
\end{equation}
Inserting the Fourier expansion of the fields in Eqs. (\ref{eq:fourier}, \ref%
{eq:fourierbar}) a little algebra yields 
\begin{equation}
\begin{split}
P_{i}={} & {\displaystyle{\displaystyle(2\pi)^{3}\sum_{p,r}}}\sum
_{r^{\prime}}\alpha(\mathbf{p})^{2}\bigl [c_{r}^{+}(\mathbf{p})c_{r^{\prime}
}(\mathbf{p})\bar{u_{r}}(\mathbf{p})\Sigma^{0\nu}p_{\nu}u_{r^{\prime} }( 
\mathbf{p})\left( 2p_{i}\right) \\
& -c_{r}^{+}(\mathbf{p})d_{r^{\prime}}^{+}(-\mathbf{p})\bar{u_{r}} (\mathbf{%
p })\Sigma^{0j}p_{j}u_{r^{\prime}}^{c}(\mathbf{-p})\left( -2p_{i}\right)
e^{2ip^{0}x_{0}} \\
& -d_{r}(\mathbf{p})c_{r^{\prime}}(-\mathbf{p})\bar{u}_{r}^{c}(\mathbf{p}
)\Sigma^{0j}p{}_{j}u_{r^{\prime}}(-\mathbf{p})\left( -2p_{i}\right)
e^{-2ip^{0}x_{0}} \\
& +d_{r}^{+}(\mathbf{p})d_{r^{\prime}}(\mathbf{p})\bar{u}_{r}^{c} (\mathbf{p}
)\Sigma^{0\nu}p_{\nu}u_{r^{\prime}}^{c}(\mathbf{p})\left( 2p_{i}\right) %
\bigr].
\end{split}%
\end{equation}
The terms appearing here are similar to the previous calculation and we
simply give the final result 
\begin{equation}
P_{i}={\displaystyle{\displaystyle\frac{(2\pi)^{3}}{V}\sum_{p,r}}}p_{i}\left[
c_{r}^{+}(\mathbf{p})c_{r}(\mathbf{p})+d_{r}^{+}(\mathbf{p})d_{r} (\mathbf{p}
)\right] .
\end{equation}

\subsection{U(1) charge}

The total current of the field is given by 
\begin{equation}
j^{\alpha}=\int d^{3}xN\left( J^{\alpha}\right) =\int d^{3}xN\left( iq\left(
(\partial_{\mu}\bar{\Psi})\Sigma^{\mu\alpha}\Psi-\bar{\Psi}
\Sigma^{\alpha\nu}(\partial_{\nu}\Psi)\right) \right) ,
\end{equation}
which after substitution of the Fourier expansion of the fields and some
manipulation yields 
\begin{equation}
\begin{split}
j^{\alpha}=q\int d^{3}xN\sum_{\mathbf{p},r}\sum_{\mathbf{p^{\prime}},r} &
i^{2}\alpha(\mathbf{p}^{\prime})\alpha(\mathbf{p})\bigl(\left[ p_{\mu
}^{\prime}+p_{\mu}\right] c_{r^{\prime}}^{+}(\mathbf{p^{\prime}} )c_{r}( 
\mathbf{p})\bar{u}_{r^{\prime}}\left( \mathbf{p}^{\prime}\right)
\Sigma^{\mu\alpha}u_{r}(\mathbf{p})e^{i\left( p^{\prime}-p\right) x} \\
& +\left[ p_{\mu}^{\prime}-p_{\mu}\right] c_{r^{\prime}}^{+} (\mathbf{\
p^{\prime}})d_{r}^{+}(\mathbf{p})u_{r}\left( \mathbf{p}^{\prime }\right)
\Sigma^{\mu\alpha}u_{r}^{c}(\mathbf{p})e^{i\left( p^{\prime }+p\right) x} \\
& -\left[ p_{\mu}^{\prime}-p_{\mu}\right] d_{r^{\prime}}(\mathbf{p}
^{\prime})c_{r}(\mathbf{p})\bar{u}_{r^{\prime}}^{c}\left( \mathbf{p}^{\prime
}\right) \Sigma^{\mu\alpha}u_{r}(\mathbf{p})e^{-i\left( p^{\prime}+p\right)
x} \\
& -\left[ p_{\mu}^{\prime}+p_{\mu}\right] d_{r^{\prime}}(\mathbf{p}
^{\prime})d_{r}^{+}(\mathbf{p})\bar{u}_{r^{\prime}}^{c}\left( \mathbf{p}
^{\prime}\right) \mathbf{\Sigma}^{\prime\mu\alpha}u_{r}^{c}(\mathbf{p}
)e^{i\left( p-p^{\prime}\right) x}\bigr).
\end{split}%
\end{equation}
For $\alpha=0$ we get the charge associated to the $U(1)$ invariance as 
\begin{equation}
\begin{split}
Q={} & \sum_{\mathbf{p},r}\sum_{r^{\prime}}\frac{i^{2}q\left( 2\pi\right)
^{3}}{2Vp_{0}}2p_{\mu}c_{r^{\prime}}^{+}(\mathbf{p})c_{r}(\mathbf{p} )\bar{
u_{r^{\prime}}}(\mathbf{p})\Sigma^{\mu0}u_{r}(\mathbf{p}) \\
& -\sum_{\mathbf{p},r}\sum_{r^{\prime}}\frac{i^{2}q\left( 2\pi\right) ^{3} }{
2Vp_{0}}2p_{\mu}d_{r}^{+}(\mathbf{p})d_{r^{\prime}}(\mathbf{p})\bar {u}
_{r^{\prime}}^{c}(\mathbf{p})\Sigma^{\mu0}u_{r}^{c}(\mathbf{p}),
\end{split}%
\end{equation}
and using again Eq. (\ref{eq:productimportant}) we get 
\begin{equation}
Q=\frac{\left( 2\pi\right) ^{3}}{V}q\sum_{\mathbf{p},r}\left( -c_{r} ^{+}( 
\mathbf{p})c_{r}(\mathbf{p})+d_{r}^{+}(\mathbf{p})d_{r}(\mathbf{p} )\right) .
\end{equation}

\subsection{Propagator}

The propagator is the expectation value of the time-ordered product of the
fields 
\begin{equation}
i\Gamma_{F}(x-y)_{ab}=\langle0|T\left( \Psi_{a}\left( x\right) \bar{\Psi }
_{b}\left( y\right) \right) |0\rangle.
\end{equation}
Substituting the Fourier expansion of the fields we get 
\begin{equation}
i\Gamma_{F}(x-y)_{ab}=\left\{ 
\begin{array}{cc}
{\displaystyle{\displaystyle\sum_{\mathbf{p}}}}\frac{1}{2V\omega_{\mathbf{p}
} }\left( \frac{S\left( \mathbf{p}\right) +m^{2}}{2m^{2}}\right)
_{ab}e^{-ip(x-y)} & x_{0}>y_{0} \\ 
{\displaystyle{\displaystyle\sum_{\mathbf{p}}}}\frac{1}{2V\omega_{\mathbf{p}
} }\left( \frac{S\left( \mathbf{p}\right) +m^{2}}{2m^{2}}\right)
_{ab}e^{ip(x-y)} & y_{0}>x_{0}%
\end{array}
\right. ,
\end{equation}
where $\omega_{\mathbf{p}}=\sqrt{\mathbf{p}^{2}+m^{2}} $ and we used the
polarization sum relations in Eq. (\ref{eq:espinoresproyector} ). We can
rewrite this equation with the help of the step function and in the
continuum limit as%
\begin{align}
i\Gamma_{F}(x-y) & =\theta\left( x_{0}-y_{0}\right) \int\frac {d^{3}\mathbf{%
p }}{\left( 2\pi\right) ^{3}2\omega_{\mathbf{p}}}\left( \frac{S\left( 
\mathbf{p }\right) +m^{2}}{2m^{2}}\right) e^{-ip(x-y)}  \notag \\
& + \theta\left( y_{0}-x_{0}\right) \int\frac{d^{3}\mathbf{p}}{\left(
2\pi\right) ^{3}2\omega_{\mathbf{p}}}\left( \frac{S\left( \mathbf{p} \right)
+m^{2}}{2m^{2}}\right) e^{ip(x-y)}.  \label{QFT_Propagator}
\end{align}

Writing $\Gamma_{F}(x-y)$ in a four-dimensional integral representation we
expect to connect with the classical Green's function, $G\left( x-y\right) $%
, obtained solving the wave equation in Eq.(\ref{eomgood}) in the presence
of sources. The Fourier transform of the Green's function, $\tilde{G}\left(
p\right) $, satisfies 
\begin{equation}
\left( \Sigma^{\mu\nu}p_{\mu}p_{\nu}-m^{2}\right) \tilde{G}\left( p\right)
=I.
\end{equation}
Using 
\begin{equation}
\left[ S\left( p\right) \right] ^{2}=p^{4},
\end{equation}
it is easy to show that%
\begin{equation}
\tilde{G}\left( p\right) =\frac{\Delta\left( p\right) }{p^{2}
-m^{2}+i\epsilon}
\end{equation}
where 
\begin{equation}
\Delta\left( p\right) =\frac{S\left( p\right) -p^{2}+2m^{2}}{2m^{2}}.
\label{Delta}
\end{equation}
Notice that we are distinguishing $S\left( \mathbf{p}\right) $ from $S\left(
p\right) $ here. We use $S\left( \mathbf{p}\right) $ when the momentum $p$
is on-shell whereas if $p$ is off-shell as in Eq. (\ref{Delta}) we use $%
S\left( p\right) $. Also notice that on-shell 
\begin{equation}
\Delta(p)|_{p^{2}=m^{2}}= \frac{S\left( \mathbf{p}\right) +m^{2}}{2m^{2}}= {%
\ \displaystyle{\displaystyle\sum_{r}}}u_{r}(\mathbf{p})\bar{u}_{r} (\mathbf{%
p} ).  \label{Deltaos}
\end{equation}
This result suggests that the appropriate four dimensional integral
representation of the two-point QFT Green's function in Eq. (\ref%
{QFT_Propagator}) is not just the direct generalization of the polarization
sum; rather, it can incorporate terms proportional to $p^{2}-m^{2}$.

In coordinates space the classical Green's function reads 
\begin{equation}
iG\left( x-y\right) =i\int\frac{d^{4}p}{\left( 2\pi\right) ^{4}} \frac{
\Delta\left( p\right) }{p^{2}-m^{2}+i\epsilon} e^{-ip\left( x-y\right) }.
\label{Green}
\end{equation}
In order to connect with (\ref{QFT_Propagator}) it is convenient to write
the above equation as%
\begin{equation}
iG\left( x-y\right) =i\int\frac{d^{3}\mathbf{p}}{\left( 2\pi\right) ^{3}}
I\left( \mathbf{p}\right) e^{i\mathbf{p}\left( \mathbf{x}-\mathbf{y} \right)
}
\end{equation}
where $I\left( \mathbf{p}\right) $ is the is the integral with respect to $%
p^{0}=\omega$%
\begin{equation}
I\left( \mathbf{p}\right) =\frac{1}{2\pi}\int_{-\infty}^{\infty}\frac {
\Delta\left( \omega,\mathbf{p}\right) e^{-i\omega\left( x-y\right) ^{0}} }{
\omega^{2}-\mathbf{p}^{2}-m^{2}+i\epsilon}d\omega.  \label{I(p)}
\end{equation}
This integral can be solved by the residue theorem in the conventional way.
However, since we are working with an unconventional extension off-shell of
the polarization sum and at the end we obtain additional terms, we give some
details of the calculation in the Appendix B. The final result for the
relation of the two-point correlation function in Eq.(\ref{QFT_Propagator})
and the integral in Eq. (\ref{Green}) is 
\begin{equation}
i\Gamma_{F}(x-y)=iG\left( x-y\right) +\frac{S^{00}-1}{2m^{2}}\delta
^{4}\left( x-y\right) .  \label{Prop}
\end{equation}

In conclusion, the two-point function in Eq. (\ref{QFT_Propagator}) is
non-covariant and differs from the covariant four-dimensional integral
representation in Eq. (\ref{Green}) by the term proportional to $\delta
^{4}\left( x-y\right) $. The non-covariance of the two-point correlation
function in the canonical quantization is a generic property of $s>1/2$
field theories. This point has been discussed in detail by Weinberg in \cite%
{Weinberg:1964cn} and we refer the reader to this reference for further
details. Concerning the calculation of the covariant S-matrix elements, the
conclusion there is that the correct Feynman rules are obtained just
skipping the non-covariant terms like the term proportional to $%
\delta^{4}\left( x-y\right) $ in Eq. (\ref{Prop}), i.e., in the calculations
we must use 
\begin{equation}\label{Propgood}
i\Gamma_{F}(x-y)=i\int\frac{d^{4}p}{\left( 2\pi\right) ^{4}} \frac {
\Delta\left( p\right) }{p^{2}-m^{2}+i\epsilon} e^{-ip\left( x-y\right) }.
\end{equation}
We remark that this four-dimensional integral representation of the
propagator incorporates terms proportional to $p^{2}-m^{2}$ to the naive
off-shell generalization of the polarization sum projector 
\begin{equation} \label{Deltasplit}
\Delta\left( p\right) =\frac{S\left( p\right) +m^{2}}{2m^{2}} -\frac {
p^{2}-m^{2}}{2m^{2}}. 
\end{equation}
This point is crucial when we incorporate interactions via the gauge
principle. Indeed, for the simplest case of interactions with $U(1)$
massless vector fields, the three-point function in momentum space is given
by 
\begin{equation}
\Gamma^{\mu}(p,p^{\prime})=\Sigma^{\mu\nu}(p^{\prime}+p)_{\nu}.
\end{equation}
It is easy to show that the Ward identity due to gauge invariance is
satisfied by this vertex with the propagator in Eq. (\ref{Propgood}) but not
with the propagator constructed only with the first term in Eq. (\ref%
{Deltasplit}).

Before ending this section we would like to remark that the algebraic
structure of the symmetric traceless symmetric tensor in Eqs. (\ref{CRS},\ref%
{ACRS}) is crucial in obtaining all the results presented in this section.


\section{Chiral decomposition and self-interactions}

The parity-based covariant basis in Eq. (\ref{parcov}) includes the
chirality operator $\chi $ with the properties 
\begin{equation}
\{\chi ,S^{\mu \nu }\}=0,\qquad \chi ^{2}=1,\qquad \lbrack \chi ,\mathcal{O}
]=0  \label{chirels}
\end{equation}%
with $\mathcal{O}$ denoting any other member of the covariant basis.

Chiral fields transforming in the $(1,0)$ (``right'' fields) and $(0,1)$
(``left'' fields) representations are defined as 
\begin{equation}
\psi_{R}=P_{R}\psi\quad\text{and} \quad\psi_{L}=P_{L}\psi,
\end{equation}
where the projectors onto well-defined chirality subspaces are given by 
\begin{equation}
P_{R}=\frac{1}{2}\left( 1+\chi\right) ,\qquad P_{L}=\frac{1}{2}\left(
1-\chi\right) .
\end{equation}
These operators have the following projector properties%
\begin{equation}
P_{R}+P_{L}=1,\qquad P_{R}P_{L}=0,\qquad P_{R}^{2}=P_{R},\qquad P_{L}
^{2}=P_{L},
\end{equation}
which together with the commutation relations in Eqs. (\ref{chirels}) imply 
\begin{equation}
\mathcal{O} P_{R,L}=P_{R,L}\mathcal{O}, \qquad
S^{\mu\nu}P_{R,L}=P_{L,R}S^{\mu\nu }.
\end{equation}

The Lagrangian in Eq. (\ref{freelag}) can be decomposed in terms of the
chiral fields as 
\begin{equation}
\mathcal{L}= \frac{1}{2}\left[ \overline{\psi_{R}}(i\partial)^{2}\psi_{L}+ 
\overline{\psi_{R}}S(i\partial)\psi_{R} + \overline{\psi_{L}} S(i\partial)
\psi_{L}\right] -m^{2} [\overline{\psi_{R}}\psi_{L} + \overline{\psi_{L}}
\psi_{R} ].
\end{equation}
The first term in the kinetic part couples left and right fields; hence, the
Lagrangian is not chirally symmetric in the massless limit. Spin 1 matter
fields cannot have chiral gauge interactions. Concerning possible applications to
hadron physics, it is not possible to realize chiral symmetry linearly and
our theory can be useful only with formalisms realizing chiral symmetry
nonlinearly. As for possible applications to model building for theories
beyond the standard model, the only possibilities for the interactions of
spin-one matter fields in this context are: i) vector gauge interactions
connected or not with the standard model group; ii) self-interactions.

Concerning interactions, we remark that the spin-one matter field has mass
dimension one, thus self-interactions are naively renormalizable. We can use
the covariant basis to classify all naively renormalizable terms in the
corresponding Lagrangian. These terms must be constructed from the following
operators bilinear in the field 
\begin{equation}
\overline{\psi}\psi,\quad\overline{\psi}\chi\psi, \quad\overline{\psi }S %
\indices{_\mu_\nu} \psi, \quad\overline{\psi}\chi S\indices{_\mu_\nu} \psi,
\quad\overline{\psi}M\indices{_\mu_\nu} \psi, \quad\overline{\psi}C_{\mu
\nu\alpha\beta} \psi, \quad\overline{\psi}\chi M\indices{_\mu_\nu} \psi,
\quad\overline{\psi}\chi C\indices{_\mu_\nu_\alpha_\beta} \psi.
\end{equation}
The last two bilinears arises from the contractions of the previous two with
the Levi-Civita tensor (contractions with the metric tensor vanish) which
can be rewritten in terms of the chirality operators using the relations 
\begin{equation}
\widetilde{M}\indices{_\mu_\nu}\equiv\frac{1}{2}\epsilon \indices{_\mu_\nu^
\rho^\sigma} M_{\rho\sigma} = -i \chi M\indices{_\mu_\nu}, \qquad\widetilde{%
C }\indices{_\mu_\nu_\alpha_\beta} \equiv\frac{1}{2} \epsilon %
\indices{_\mu_\nu^\rho^\sigma} C_{\rho\sigma\alpha\beta} = -i \chi C %
\indices{_\mu_\nu_\alpha_\beta}.
\end{equation}
There are ten independent non-vanishing Lorentz invariant terms that can be
built from the products of these bilinears. The most general naively
renormalizable self-interaction Lagrangian is given by 
\begin{equation}
\begin{split}
\mathcal{L}_{self} & = c_{1}\left( \overline{\psi}\psi\right) ^{2}
+c_{2}\left( \overline{\psi}\chi\psi\right) ^{2} +c_{3}\left( \overline{\psi}
S\indices{_\mu_\nu} \psi\right) ^{2} + c_{4}\left( \overline{\psi}\chi S %
\indices{_\mu_\nu} \psi\right) ^{2} \\
& + c_{5}\left( \overline{\psi}M\indices{_\mu_\nu} \psi\right) ^{2}
+c_{6}\left( \overline{\psi}C\indices{_\mu_\nu_\alpha_\beta} \psi\right)
^{2} + c_{7}\left( \overline{\psi}\psi\right) \left( \overline{\psi}
\chi\psi\right) + c_{8}\left( \overline{\psi}S\indices{_\mu_\nu} \psi\right)
\left( \overline{\psi}\chi S\indices{^\mu^\nu} \psi\right) \\
& + c_{9}\left( \overline{\psi}M\indices{_\mu_\nu} \psi\right) \left( 
\overline{\psi}\chi M\indices{^\mu^\nu} \psi\right) +c_{10}\left( \overline{
\psi}C\indices{_\mu_\nu_\alpha_\beta} \psi\right) \left( \overline{\psi}\chi
C\indices{^\mu^\nu^\alpha^\beta} \psi\right) .
\end{split}%
\end{equation}
Some of these terms violate discrete symmetries and it would be interesting
to explore the consequences of the existence of spin-one matter particles in
physics beyond the standard model, in particular if it could play a role in
resolving the dark matter enigma. If the massless limit of our formalism is
a sensible theory, all these coefficient must vanish since all these terms
violate the gauge invariance in Eqs. (\ref{gaugeinv}).


\section{Summary and conclusions}

In this work we introduce a Dirac-like formalism for the description of spin
1 massive fields transforming in the $(1,0)\oplus(0,1)$ representation of the HLG.
The formalism is based on the simultaneous projection on parity eigenspaces
and on the appropriate Poincar\'e orbit. This projection is done using the
parity-based covariant basis for the matrix operators acting on the $%
(1,0)\oplus(0,1)$ representation space constructed in \cite%
{Gomez-Avila:2013qaa}. We construct the charge conjugation operator and show
that it commutes with parity. An explicit construction of the solutions
using the representation of operators in the basis of well-defined parity
for $(1,0)\oplus(0,1)$ shows that the ``down'' component of the solutions
are suppressed as $v/c$ with respect to the ``up'' part of the solutions in
the non-relativistic limit. More importantly, the ``down'' part of the
solutions are fixed by the kinematics, as a consequence of the constrained
dynamics. We work out the constraints at the classical field theory level,
show that the system has only second class constraints, and obtain the Dirac
bracket of the canonical conjugate variables. We carry out the canonical
quantization of the theory, and calculate commutator relations for the
canonical variables consistent with the classical Dirac brackets. Sensible
results are obtained for the relevant physical quantities: energy, momentum, 
$U(1)$ charge, and the propagator. The algebraic properties of the covariant
basis are instrumental in obtaining these results. With the aid of the
chirality operator which naturally appears in the construction of the
covariant basis, we analyse the chiral structure of the theory finding 
that spin-one  matter fields cannot have chiral gauge interactions, 
but admit vector gauge interactions. Spin-one matter fields have mass-dimension one 
therefore self-interactions are naively renormalizable. Using the covariant basis, 
we classify all renormalizable self-interaction terms. 

Although the formalism is designed for massive particles, the classical theory
has a soft $m\to 0$ limit, in whose case first class constraints (gauge 
symmetries) appear. It would be interesting to explore if a sensible 
quantum field theory can be obtained in this case.


\begin{acknowledgments}
Work supported by CONACYT (M\'exico) under project CB-156618, SNI grants and
Red FAE project. We also acknowledge financial support from Universidad de
Guanajuato through a DAIP project and Secretaria de Educacion P\'ublica,
M\'exico under a PIFI project.
\end{acknowledgments}

\appendix

\section{Constrained dynamics with a $U(1)$ coupling}
\label{app:u1coupled}
The Lagrangian with $U\left( 1\right) $ coupling is%
\begin{align}
\mathcal{L}& =D_{\mu }\bar{\Psi}\Sigma ^{\mu \nu }D_{\nu }\Psi -m^{2}\bar{%
\Psi}\Psi , \\
D_{\mu }\Psi & =\partial _{\mu }\Psi + i q A_{\mu }\Psi , \\
D_{\mu }^{\dagger }\bar{\Psi}& =\partial _{\mu }\bar{\Psi} -i q A_{\mu }\bar{%
\Psi}.
\end{align}%
In terms of the \textquotedblleft up\textquotedblright\ $\varphi $ and
\textquotedblleft down\textquotedblright\ $\xi $ fields, the Lagrangian reads
\begin{align}
\mathcal{L}& =D_{0}^{\dagger }\varphi ^{\dagger }D_{0}\varphi -\frac{1}{2}%
\left( D_{0}^{\dagger }\varphi ^{\dagger }D_{i}J^{i}\xi +D_{0}^{\dagger }\xi
^{\dagger }D_{i}J^{i}\varphi \right) -\frac{1}{2}\left( D_{i}^{\dagger
}\varphi ^{\dagger }J^{i}D_{0}\xi +D_{i}^{\dagger }\xi ^{\dagger
}J^{i}D_{0}\varphi \right)  \notag \\
& +D_{i}^{\dagger }\varphi ^{\dagger }D^{i}\varphi +\frac{1}{2}%
D_{i}^{\dagger }\varphi ^{\dagger }\left\{ J^{i},J^{j}\right\} D_{j}\varphi +%
\frac{1}{2}D_{i}^{\dagger }\xi ^{\dagger }\left\{ J^{i},J^{j}\right\}
D_{j}\xi -m^{2}\left( \varphi ^{\dagger }\varphi -\xi ^{\dagger }\xi \right)
,
\end{align}%

For the purposes of quantization, it is instructive to analyse the
Dirac bracket. The canonical momenta in the presence of a $U(1)$
coupling are%
\begin{align}
\pi _{a}& =\frac{\delta \mathcal{L}}{\delta \left( \partial _{0}\varphi
_{a}\right) }=D_{0}^{\dagger }\varphi _{a}^{\dagger }-\frac{1}{2}\left(
D_{i}^{\dagger }\xi ^{\dagger }J^{i}\right) _{a}, \\
\pi _{a}^{\dagger }& =\frac{\delta \mathcal{L}}{\delta \left( \partial
_{0}\varphi _{a}^{\dagger }\right) }=D_{0}\varphi _{a}-\frac{1}{2}\left(
J^{i}D_{i}\xi \right) _{a} , \label{Momento_psi_bar}
\end{align}%
and%
\begin{align}
\tau _{a}& =\frac{\delta \mathcal{L}}{\delta \left( \partial _{0}\xi
_{a}\right) }=-\frac{1}{2}\left( D_{i}\varphi ^{\dagger }J^{i}\right) _{a},
\label{Momento_phi} \\
\tau _{a}^{\dagger }& =\frac{\delta \mathcal{L}}{\delta \left( \partial
_{0}\xi _{a}^{\dagger }\right) }=-\frac{1}{2}\left( J^{i}D_{i}\varphi
\right) _{a},  \label{Momento_phi_bar}
\end{align}%
which imply constraints even in the presence of electromagnetic interactions.

The Hamiltonian density, incorporating the constraints as Lagrange
multipliers, is%
\begin{align}
\mathcal{H}^{\ast }& =\pi _{a}\pi _{a}^{\dagger }+\frac{1}{2}\pi _{a}\left(
J^{i}D_{i}\xi \right) _{a}+\frac{1}{2}\left( D_{i}^{\dagger }\xi ^{\dagger
}J^{i}\right) _{a}\pi _{a}^{\dagger }+\frac{1}{4}\left( D_{i}^{\dagger }\xi
^{\dagger }J^{i}\right) _{a}\left( J^{j}D_{j}\xi \right) _{a}  \notag \\
& -D_{i}^{\dagger }\varphi ^{\dagger }D^{i}\varphi -\frac{1}{2}%
D_{i}^{\dagger }\varphi ^{\dagger }\left\{ J^{i},J^{j}\right\} D_{j}\varphi -%
\frac{1}{2} D_{i}^{\dagger }\xi ^{\dagger }\left\{ J^{i},J^{j}\right\}
D_{j}\xi +m^{2}\left( \varphi ^{\dagger }\varphi -\xi ^{\dagger }\xi \right)
\notag \\
& +ieA_{0}\left[ \pi _{a}\varphi _{a}-\varphi _{a}^{\dagger }\pi
_{a}^{\dagger } \right] +\frac{ie}{2}A_{0}\left[ \xi _{a}^{\dagger }\left(
J^{i}D_{i}\varphi \right) _{a}-\left( D_{i}^{\dagger }\varphi ^{\dagger
}J^{i}\right) _{a}\xi _{a} \right]  \notag \\
& +\lambda _{a}\rho _{a}+\mathbb{\lambda }_{a}^{\dagger }\rho _{a}^{\dagger
},
\end{align}%
and the Hamilton equations are%
\begin{align}
\partial _{0}\varphi _{a}& =\frac{\delta H^{\ast }}{\delta \pi _{a}}=\pi
_{a}^{\dagger }+\frac{1}{2}\left( J^{i}D_{i}\xi \right) _{a} +i e A_0
\varphi_a, \\
\partial _{0}\pi _{a}^{\dagger }& =-\frac{\delta H^{\ast }}{\delta \varphi
_{a}^{\dagger }}=-D_{i}D^{i}\varphi _{a}-\frac{1}{2}D_{i}D_{j}\left( \left\{
J^{i},J^{j}\right\} \varphi \right) _{a}-m^{2}\varphi _{a}+\frac{1}{2}\left(
J^{i}D_{i}^{\dagger }\lambda \right) _{a} \nonumber \\
& +ie\left[ A_{0}\pi ^{\dagger }-\frac{1}{2}\left( J^{i}D_{i}A_{0}\xi
\right) \right] _{a} \\
\partial _{0}\xi _{a}& =\frac{\delta H^{\ast }}{\delta \tau _{a}}=\lambda
_{a}, \\
\partial _{0}\tau _{a}^{\dagger }& =-\frac{\delta H^{\ast }}{\delta \xi
_{a}^{\dagger }}=\frac{1}{2}D_{i}\left( J^{i}\pi ^{\dagger }\right) _{a}+ 
\frac{1}{4}\left( J^{i}J^{j}D_{i}D_{j}\xi \right) _{a}-\frac{1}{2}
D_{i}D_{j}\left( \left\{ J^{i},J^{j}\right\} \xi \right) _{a} \nonumber \\
& +m^{2}\xi _{a}-\frac{ie}{2}A_{0}\left( J^{i}D_{i}\varphi \right) _{a}.
\end{align}%

The temporal evolution of any dependent dynamic variable fields and momenta
can be written as%
\begin{equation}
\dot{B}=\frac{\partial B}{\partial t}+\left\{ B,H^{\ast }\right\},
\end{equation}%
and again they have the same Poisson brackets between fields and canonical
momenta, Eqs (\ref{Poisson_B_Fields}).

The dynamics generated by the modified Hamiltonian must preserve the
restrictions%
\begin{equation}
\partial _{0}\rho _{a}=\frac{\partial \rho _{a}}{\partial t}+\left\{ \rho
_{a},H^{\ast }\right\} =0.
\end{equation}%
This leads to secondary constraints%
\begin{equation}
\kappa _{a}=D_{i}^{\dagger }\left( \pi J^{i}\right) _{a}-\frac{1}{2}\left(
D_{j}^{\dagger }D_{i}^{\dagger }\xi ^{\dagger }J^{j}J^{i}\right)
_{a}+m^{2}\xi _{a}^{\dagger }+\frac{ie}{2}F_{0i}\left( \varphi ^{\dagger
}J^{i}\right) _{a}=0.
\end{equation}%
and, again, for consistency, requires 
\begin{equation}
\partial _{0}\kappa _{a}=\frac{\partial \kappa _{a}}{\partial t}+\left\{
\kappa _{a},H^{\ast }\right\} =0.
\end{equation}%
This condition yields%
\begin{align}
\dot{\kappa}& =-D_{k}^{\dagger }\left[ D_{j}^{\dagger }D^{\dagger j}\varphi +%
\frac{1}{2}D_{j}^{\dagger }D_{i}^{\dagger }\left( \varphi ^{\dagger }\left\{
J^{i},J^{j}\right\} \right) \right] J^{k}-m^{2}D_{k}^{\dagger }\varphi
^{\dagger }J^{k} \nonumber \\
& +\lambda ^{\dagger }\left( \frac{ie}{2}F_{ki}J^{k}J^{i}+m^{2}\right)
-ieD_{i}^{\dagger }\left[ A_{0}\pi -\frac{1}{2}\left( D_{j}^{\dagger }\left(
A_{0}\xi ^{\dagger }\right) J^{j}\right) \right] J^{i} \nonumber \\
& +\frac{ie}{2}F_{0k}\left( \pi -ieA_{0}\varphi ^{\dagger }+\frac{1}{2}%
\left( D_{i}^{\dagger }\xi ^{\dagger }J^{i}\right) \right) J^{k}+\frac{ie}{2}%
\dot{F}_{0i}\varphi ^{\dagger }J^{i} \nonumber \\
& -\frac{ie}{2}\left( \dot{A}_{j}D_{i}^{\dagger }\xi ^{\dagger
}J^{j}J^{i}+D_{j}^{\dagger }\dot{A}_{i}\xi ^{\dagger }J^{j}J^{i}\right) +ie%
\dot{A}_{i}\pi J^{i} = 0
\end{align}%
Although this is a complicated equation, it just defines
$\lambda^{\dagger }$ and does not give rise to additional secondary
constraints.

Now we write the Poisson brackets between the constraints. It is
straightforward to see that%
\begin{align}
\left\{ \rho _{a}\left( \mathbf{x}\right) ,\rho _{b}\left( \mathbf{y}\right)
\right\} & =0, \\
\left\{ \rho _{a}\left( \mathbf{x}\right) ,\rho _{b}^{\dagger }\left( 
\mathbf{y}\right) \right\} & =0, \\
\left\{ \rho _{a}\left( \mathbf{x}\right) ,\kappa _{b}\left( \mathbf{y}%
\right) \right\} & =0.
\end{align}%
However, for $\left\{ \rho _{a}\left( \mathbf{x}\right) ,\chi _{b}^{\dagger
}\left( \mathbf{y}\right) \right\} $ we get 
\begin{align}
\left\{ \rho _{a}\left( \mathbf{x}\right) ,\kappa _{b}^{\dagger }\left( 
\mathbf{y}\right) \right\} & =\left\{ \left[ \tau _{a}+\frac{1}{2}\left(
D_{k}^{\dagger }\varphi ^{\dagger }J^{k}\right) _{a}\right] \left( \mathbf{x}%
\right) ,\left[ D_{i}\left( J^{i}\pi ^{\dagger }\right) _{b}-\frac{1}{2}%
\left( D_{i}D_{j}J^{i}J^{j}\xi \right) _{b}+m^{2}\xi _{b}\right] \left( 
\mathbf{y}\right) \right\} \nonumber \\
& -\frac{ie}{2}\left\{ \left[ \tau _{a}+\frac{1}{2}\left( D_{k}^{\dagger
}\varphi ^{\dagger }J^{k}\right) _{a}\right] \left( \mathbf{x}\right) ,\left[
F_{0i}\left( J^{i}\varphi \right) _{a}\right] \left( \mathbf{y}\right)
\right\} ,
\end{align}%
and since the last line of this equation vanishes,%
\begin{align}
\left\{ \rho _{a}\left( \mathbf{x}\right) ,\kappa _{b}^{\dagger }\left( 
\mathbf{y}\right) \right\} & =-\frac{1}{2}\left\{ \tau _{a}\left( \mathbf{x}%
\right) ,\left( D_{i}D_{j}J^{i}J^{j}\xi \right) _{b}\left( \mathbf{y}\right)
\right\} +m^{2}\left\{ \tau _{a}\left( \mathbf{x}\right) ,\xi _{b}\left( 
\mathbf{y}\right) \right\} \nonumber \\
& +\frac{1}{2}\left\{ \left( D_{k}^{\dagger }\varphi ^{\dagger }J^{k}\right)
_{a}\left( \mathbf{x}\right) ,D_{i}\left( J^{i}\pi ^{\dagger }\right)
_{b}\left( \mathbf{y}\right) \right\} .
\end{align}%
This can be written as%
\begin{equation}
\left\{ \rho _{a}\left( \mathbf{x}\right) ,\kappa _{b}^{\dagger }\left( 
\mathbf{y}\right) \right\} =\frac{1}{2}\left[ D_{\mathbf{y}i}D_{\mathbf{y}%
j}+D_{\mathbf{y}i}D_{\mathbf{x}j}^{\dagger }\right] \delta ^{3}\left( 
\mathbf{x-y}\right) \left( J^{j}J^{i}\right) _{ba}-m^{2}\delta _{ab}\delta
^{3}\left( \mathbf{x-y}\right) .
\end{equation}%
In this expression, we can change $\partial _{x}$ by $-\partial _{y}$ and $%
A_{j}\left( \mathbf{x}\right) $ by $A_{j}\left( \mathbf{y}\right) $ in $D_{%
\mathbf{x}j}^{\dagger }\delta ^{3}\left( \mathbf{x-y}\right) $ to get
$-D_{\mathbf{y}j}\delta ^{3}\left( \mathbf{x-y}\right) $. This
allows us to conclude that 
\begin{equation}
\left\{ \rho _{a}\left( \mathbf{x}\right) ,\kappa _{b}^{\dagger }\left( 
\mathbf{y}\right) \right\} =-m^{2}\delta _{ab}\delta ^{3}\left( \mathbf{x-y}%
\right) .
\end{equation}%
These are equal to the free field Poisson brackets.

\section{Integral representation of the propagator}

For the calculation of the integral in Eq. (\ref{I(p)}) we split
it into the real axis and the semi-circle contributions 
\begin{equation}
\frac{1}{2\pi}\oint_{C}\frac{\Delta\left( \omega,\mathbf{p}\right)
e^{-i\omega\left( x-y\right) ^{0}}}{\omega^{2}-\mathbf{p}^{2}-m^{2}
+i\epsilon}d\omega=I\left( \mathbf{p}\right) +\frac{1}{2\pi}\int_{C_{R} } 
\frac{\Delta\left( \omega,\mathbf{p}\right) e^{-i\omega\left( x-y\right)
^{0}}}{\omega^{2}-\mathbf{p}^{2}-m^{2}+i\epsilon}d\omega.
\end{equation}
Causality requires us to close the contour $C$ with a (counterclockwise)
semicircle on the upper complex plane for $\left( x-y\right) ^{0}<0$ and
with a (clockwise) semicircle on the lower plane for $\left( x-y\right)
^{0}>0$. In the case $\left( x-y\right) ^{0}>0,$ $C$ encloses the pole $%
\omega_{\epsilon}=\sqrt{\mathbf{p}^{2}+m^{2}-i\epsilon}$ and we get 
\begin{equation}
I\left( \mathbf{p}\right) =\frac{-i\Delta\left( \omega_{\epsilon },\mathbf{p}
\right) e^{-i\omega_{\epsilon}\left( x-y\right) ^{0}}} {2\omega_{\epsilon}}- 
\frac{1}{2\pi}\int_{C_{R}^{-}}\frac{\Delta\left( \omega,\mathbf{p}\right)
e^{-i\omega\left( x-y\right) ^{0}}}{\omega ^{2}-\mathbf{p}
^{2}-m^{2}+i\epsilon}d\omega.  \label{Integral_Cr1}
\end{equation}
Similarly, for $\left( x-y\right) ^{0}<0$ we obtain 
\begin{equation}
I\left( \mathbf{p}\right) =\frac{-i\Delta\left( -\omega_{\epsilon },\mathbf{%
p }\right) e^{i\omega_{\epsilon}\left( x-y\right) ^{0}}} {2\omega_{\epsilon}}%
- \frac{1}{2\pi}\int_{C_{R}^{+}}\frac{\Delta\left( \omega,\mathbf{p}\right)
e^{-i\omega\left( x-y\right) ^{0}}}{\omega ^{2}-\mathbf{p}
^{2}-m^{2}+i\epsilon}d\omega.  \label{Integral_Cr2}
\end{equation}
Next we parameterize $\omega\in C_{R}^{\pm}$ as $\omega=Re^{i\theta}$ with $%
0\leq\theta\leq\pi$ for $C_{R}^{+}$ and $\pi\leq\theta\leq2\pi$ for $%
C_{R}^{-}$. For large $R$ we get 
\begin{equation}
\lim_{R\rightarrow\infty}\frac{\Delta\left( Re^{i\theta},\mathbf{p}\right) }{
R^{2}e^{2i\theta}-\mathbf{p}^{2}-m^{2}+i\epsilon}=\frac{1}{2m^{2}}\left(
S^{00}-1\right) \neq0 ,
\end{equation}
and unlike the scalar and fermion case, the integrals over $C_{R}^{\pm}$ do
not vanish 
\begin{equation}
\lim_{R\rightarrow\infty}\int_{C_{R}^{\pm}}\frac{\Delta\left( \omega , 
\mathbf{p}\right) e^{-i\omega\left( x-y\right) ^{0}}}{\omega ^{2}-\mathbf{p}
^{2}-m^{2}+i\epsilon}d\omega=\frac{\left( S^{00}-1\right) }{2m^{2}}
\int_{C_{R}^{\pm}}e^{-i\omega\left( x-y\right) ^{0}}d\omega.  \label{ICR1}
\end{equation}
The integral on the r.h.s of Eq.(\ref{ICR1}) is readily obtained as 
\begin{equation}
\int_{C_{R}^{\pm}}e^{-i\omega\left( x-y\right) ^{0}}d\omega=-2\pi
\delta\left( x^{0}-y^{0}\right) .  \label{ICR2}
\end{equation}
Using Eqs. (\ref{ICR1}, \ref{ICR2}) we can rewrite Eqs. (\ref{Integral_Cr1},%
\ref{Integral_Cr2}) as 
\begin{align}
I\left( \mathbf{p}\right) & =\frac{-i\Delta\left( \omega_{\epsilon },\mathbf{%
\ p}\right) e^{-i\omega_{\epsilon}\left( x-y\right) ^{0}}} {%
2\omega_{\epsilon}} +\frac{\left( S^{00}-1\right) }{2m^{2}}\delta\left(
x^{0}-y^{0}\right) ;\qquad\left( x-y\right) ^{0}>0, \\
I\left( \mathbf{p}\right) & =\frac{-i\Delta\left( -\omega_{\epsilon }, 
\mathbf{p}\right) e^{i\omega_{\epsilon}\left( x-y\right) ^{0}}} {
2\omega_{\epsilon}}+\frac{\left( S^{00}-1\right) }{2m^{2}}\delta\left(
x^{0}-y^{0}\right) ;\qquad\left( x-y\right) ^{0}<0,
\end{align}
and Eq.(\ref{Green}) reads 
\begin{align}
iG\left( x-y\right) & =\frac{\theta\left( x^{0}-y^{0}\right) }{\left(
2\pi\right) ^{3}}\int\frac{d^{3}\mathbf{p}}{2\omega_{\epsilon}}\Delta\left(
\omega_{\epsilon},\mathbf{p}\right) e^{-i\omega\left( x-y\right) ^{0} }e^{i 
\mathbf{p}\left( \mathbf{x}-\mathbf{y}\right) }  \notag \\
& +\frac{\theta\left( y^{0}-x^{0}\right) }{\left( 2\pi\right) ^{3}} \int 
\frac{d^{3}\mathbf{p}}{2\omega_{\epsilon}}\Delta\left( -\omega_{\epsilon }, 
\mathbf{p}\right) e^{+i\omega\left( x-y\right) ^{0}}e^{i\mathbf{p}\left( 
\mathbf{x}-\mathbf{y}\right) }  \notag \\
& +\frac{S^{00}-1}{2m^{2}}\delta^{4}\left( x-y\right) .  \label{G3}
\end{align}
Changing $\mathbf{p}$ by $-\mathbf{p}$ in the second line of Eq. (\ref{G3}),
taking the $\epsilon\rightarrow0$ limit and using%
\begin{equation}
\Delta\left( \omega_{\mathbf{p}},p\right) =\frac{S\left( \mathbf{p}\right)
+m^{2}}{2m^{2}},
\end{equation}
we finally obtain 
\begin{equation}
i\Gamma_{F}(x-y)=iG\left( x-y\right) +\frac{S^{00}-1}{2m^{2}}\delta
^{4}\left( x-y\right) .  \label{Propagator}
\end{equation}

\bibliographystyle{prsty}
\bibliography{highspin}

\end{document}